\documentclass[11pt]{article}
\usepackage{graphicx,color}
\usepackage[utf8]{inputenc}
\usepackage{a4,amssymb,amsmath}

\def\sig{\sigma} \def\kap{\kappa} \def\lam{\lambda}
\def\Tr{\mathrm{Tr}\,}
\def \RR {{\mathbb R}}
\def \CC {{\mathbb C}}

\def\SO{\mathrm{SO}}
\def\Eu{\mathrm{E}}
\def \merw#1#2{{}_{#1}M^{#2}}
 
\def\lra{\leftrightarrow}
\def\AP{A^{\rm P}} \def\AK{A^{\rm F}} \def\FP{F^{\rm P}} 
\def\fz{{(\mathrm{Fierz})}}
\def\red{\mathrm{red}}
\def\tpf{2-point function}
\def\set{stress-energy tensor}
\def\inv{^{-1}}
\def\wh{\widehat} \def\wt{\widetilde} 
 
\def\sumno{\sum\nolimits}
\def\prodno{\prod\nolimits}
\def\qed{\hfill$\square$\medskip}

\def\bpm{\begin{pmatrix}} \def\epm{\end{pmatrix}}

\def\be{\begin{equation}} \def\bea{\begin{eqnarray}}
\def\ba{\begin{array}} \def\eea{\end{eqnarray}} 
\def\ee{\end{equation}}\def\ea{\end{array}}

\newcounter{allcount} \numberwithin{allcount}{section}

\newtheorem{lemma}[allcount]{Lemma}
\newtheorem{propo}[allcount]{Proposition}
\newtheorem{coro}[allcount]{Corollary}
\newtheorem{remark}[allcount]{Remark}
\newtheorem{example}[allcount]{Example}

\newcommand{\eref}[1]{Eq.~(\ref{#1})}

\newcommand{\cref}[1]{Cor.~\ref{#1}}
\newcommand{\lref}[1]{Lemma~\ref{#1}}
\newcommand{\pref}[1]{Prop.~\ref{#1}}
\newcommand{\rref}[1]{Remark~\ref{#1}}
\newcommand{\xref}[1]{Example~\ref{#1}}
\newcommand{\sref}[1]{Sect.~\ref{#1}}
\newcommand{\aref}[1]{App.~\ref{#1}}

\begin{document}

\title{\vskip-10mm Helicity decoupling \\ in the massless limit of massive tensor
  fields\footnote{An abridged version of this paper, focusing on spin
    $s=1$ and $s=2$, is \cite{MRS}.}}

\author{{\sc Jens Mund} 
\\ \small 
Departamento de F\'isica, Universidade Federal de Juiz de Fora,
\\ \small
Juiz de Fora 36036-900, MG, Brazil 
\\ \small
email: mund@fisica.ufjf.br \\ \phantom{X} \\
{\sc Karl-Henning Rehren} 
\\ \small
 Institut f\"ur Theoretische Physik, Universit\"at G\"ottingen, 37077 G\"ottingen,
 Germany 
\\ \small
email: rehren@theorie.physik.uni-goettingen.de \\
\phantom{X} \\
{\sc Bert Schroer} 
\\ \small
Institut f\"ur Theoretische Physik der FU Berlin, Berlin, Germany 
\\ \small
Centro Brasileiro de Pesquisas F\'isicas, 22290-180 Rio de Janeiro,
RJ, Brazil 
\\ \small
email: schroer@zedat.fu-berlin.de}

\maketitle

\begin{abstract} Massive and massless potentials play an essential
  role in the perturbative formulation of particle interactions. Many
  difficulties arise due to the indefinite metric in gauge theoretic
  approaches, or the increase with the spin of the UV dimension of
  massive potentials. All these problems can be evaded in one stroke:
  modify the potentials by suitable terms that leave unchanged the field
  strengths, but are not polynomial in the momenta. This feature
  implies a weaker localization property: the potentials are
  ``string-localized''. In this setting, several old issues can be
  solved directly in the physical Hilbert space of the respective
  particles: We can control the separation of helicities in the
  massless limit of higher spin fields and conversely we recover
  massive potentials with $2s+1$ degrees of freedom by a smooth
  deformation of the massless potentials (``fattening''). We construct
  \set s for massless fields of any helicity (thus
  evading the Weinberg-Witten theorem). We arrive at a simple
  understanding of the van Dam-Veltman-Zakharov discontinuity
  concerning, e.g., the distinction between a massless or a very light
  graviton. Finally, the use of string-localized fields opens new
  perspectives for interacting quantum field theories with, e.g.,
  vector bosons or gravitons.   
\end{abstract} 

 \vskip20mm

\section{Overview}
\setcounter{equation}{0}
\label{s:overview}

The purpose of this contribution is to formulate and investigate a
unified setting for potentials describing both massless and massive
vector and tensor bosons, that live in Hilbert space (i.e., without
negative-norm states even at intermediate steps.) 
The Hilbert space is that of the field strengths, which have no positivity
problems. Our focus is here on the free fields, and in particular on
the limit $m\to0$ that is smooth in this setting. We comment in
appropriate places on the issues concerning renormalizable
interactions, in particular the improved UV behaviour, and refer for
more details to the literature \cite{S15,S16,MO,GMV,MS,M}.

Hilbert space positivity is quantum theory's most basic attribute
which is indispensible for its probability interpretation. In the standard
formulation, it enters through the identification of quantum states
with unit rays in a Hilbert space. The von Neumann uniqueness theorem
concerning Heisenberg's commutation relations in terms of operators in
a Hilbert space secures the positivity of quantum mechanics. With the
more general notion of states as positive linear 
functionals of *-algebras, positivity is secured via the GNS
construction leading again to representations in Hilbert space.

In contrast to Born's quantum mechanical localization in terms of
probability (the argument of the wave function directly refers to the
position of a particle), the positivity issue in QFT is more demanding
and detaches the causal localization of fields from the localization
of particles: it is impossible to assign a probability such that
$\phi(x)$ creates with certainty a particle at the point $x$ \cite{NW}. 
Classical field theory has no structural feature whose quantization
guarantees that the corresponding  quantum fields act in a Hilbert
space, and canonical quantization of massless fields generically
introduces negative-norm states. For low
spin $s<1$ this problem is absent but, as Gupta and Bleuler first
pointed out, starting from $s=1$, only by using additional
negative-metric degrees of 
freedom can one maintain a formal analogy to the classical covariant
gauge potentials at a point.  

There has been extensive work on higher spin fields by Fronsdal 
\cite{Fd}, Rindani et al.\ \cite{RS}, Francia et al.\ \cite{FMS},
Bekaert et al.\ \cite{BBS}, Sagnotti \cite{S13}, and in
particular by Vasiliev \cite{FV,V1,V2,DS14,RT15} 
(to name only a few)\footnote{The strings in the titles
  of some of these papers have nothing in common with the strings
  in our paper. The prevalent idea to assign a ``quantum mechanical''
  notion of localization to (super)strings is at variance with the causal 
  localization of quantum field theory, as discussed in
  \cite[Sect.~2]{S14}.}. While Fronsdal proves the  
positivity of the \tpf\ contracted with constrained sources, most of
the more recent work concentrates on Lagrangians and field equations
without even addressing the crucial issues of quantization: positivity
(semi-definite two-point functions, actions of fields in a ghost-free
Hilbert space, \dots) and causal localization (commutation relations at
spacelike distance).

An alternative to canonical quantization is to start from Wigner's 
classification of unitary positive energy representations of the
Poincar\'e group. This approach takes care of positivity from the
start, and one can fully concentrate on the interplay of covariance
with causal localizability. In order to construct covariant fields on
the Wigner Fock space, one uses intertwiner functions $u_\alpha(p)$ 
which mediate between unitary representations of the stabilizer group
and matrix representations for covariant field multiplets; for $m>0$
this was done by Weinberg \cite{W}. Since the intertwiners determine the
two-point functions, they also determine the commutators, and
therefore control causal commutativity. In the case of half-integer
spin, causality is incompatible with positivity and has to be replaced
by anti-commutation relations instead. For spin 1, as expected,
massless vector potentials $A_\mu(x)$ localized at points $x$ (which
are precisely what is needed for QED) do not exist on the Hilbert space. 

During the last decade we have learned that the Wigner representation
theory contains much more information concerning causal localization and
associated intertwiners \cite{MSY,PY}. Modifying the intertwiners does
not change the particle content, but will modify properties of the fields. 
It may improve the short-distance dimension at the price of a
weaker localization. With the new flexibility, one can construct
massive vector potentials that admit renormalizable interactions and 
massless vector potentials directly on the Hilbert space, and even
causal fields that transform in Wigner's infinite-spin representation
\cite{MSY}. All these fields are localized on ``strings'' = rays
extending from a point to spacelike infinity. 

This kind of localization was actually proven long ago to be
{\em necessary} to connect the vacuum with charged one-particle
states. Buchholz \cite{Bu} gave a nonperturbative proof that
electrical charge-carrying quantum fields cannot be compactly
localized and that the tightest localized covariant fields cannot be
better localized than on spacelike half-lines. Buchholz and
Fredenhagen \cite{BF}, by an analysis of how the corresponding state
functionals on the observables differ, showed that string-localization
may be necessary in massive theories; and at the same time {\em
  sufficient} to construct many-particle scattering states in the
asymptotic time limits.  

Taking advantage of this new flexibility, one can reformulate all
perturbative interactions of the Standard Model directly on the
Hilbert space of the physical asymptotic particles, without recourse
to ghosts and BRST methods. The emerging program of causal
perturbation theory with string-localized fields \cite{S15,S16,MO,GMV,MS,M}
suggests that it allows to compute the 
same scattering matrix as the BRST gauge theory approach, but unlike
the latter, it also allows to construct (non-observable) interpolating
fields between the vacuum and the charged states, that live on the
Hilbert space and are just local enough to allow for scattering theory. 

The reformulation is classically
equivalent to the usual one in that the Lagrangian differs only by a
total derivative. 

\begin{example}\label{x:QED} \cite{S16}
We want to illustrate this in the case of massive or
massless vector bosons coupled to a conserved current $j^\mu(x)$. The
interaction part of the action is
\bea
S_{\rm int} = \int d^4x\, A_\mu(x)j^\mu(x).
\eea
Unlike the integrand, the integral does not depend on the choice of
the potential when the field strength is given. Namely, by Poincar\'e's
Lemma, any two potentials differ by a gradient $\partial_\mu\alpha$,
and $\int d^4x\, \partial_\mu\alpha j^\mu = \int
d^4x\, \partial_\mu(\alpha j^\mu) = 0$. Indeed, the classical equation of
motion derived by Hamilton's Principle contains only the field
strength. 

We are going to use potentials that depend on a direction $e$ in
Minkowski spacetime and are manifestly localized along the string
$x+\RR^+_{0}e$:
\bea\label{AFe}
A_\mu(x,e) = \int\nolimits_0^\infty ds \, F_{\mu\nu}(x+se)e^\nu.
\eea
One easily sees that $\partial_\mu A_\nu(x,e)-\partial_\nu A_\mu(x,e)
= F_{\mu\nu}(x)$. From these facts, the $e$-independence of
\bea\label{Se1}
S_{\rm int}(e) = \int d^4x\, A_\mu(x,e)j^\mu(x)
\eea
is manifest.
\end{example}

The preservation of the equivalence of $S_{\rm int}(e)$ and $S_{\rm int}$
at the quantum level, i.e., the $e$-independence of the S matrix
$T\exp i\int d^4x\, L_{\rm int}(e)$, is an issue of renormalization conditions,
whose satisfyability beyond the lowest orders is presently under
investigation. Interacting fields can then be constructed with the
Bogoliubov S matrix method \cite{Bg}. The interacting potential and
Dirac field will depend on $e$, but their field strength and current
are expected to be $e$-independent. 

The merit of the string-localized approach is twofold:
  String-localized interacting fields that connect the vacuum with
  scattering states are constructed in the Hilbert space, and in such a way
  that observables are string-independent, and hence causality
  is secured. In the massless case,  
$A_\mu(x,e)$ is defined on the physical Hilbert space of the field
strength which has only two photon states; and in the massive case, 
$A_\mu(x,e)$ has UV dimension $d_{UV}=1$ instead of $d_{UV}=2$
for the Proca potential $\AP_\mu$, see \eref{AA1} vs.\
\eref{APAP}. Thus, the coupling $A_\mu(x,e)j^\mu(x)$ of a massive
vector boson to the Dirac current of dimension $d_{UV}=3$ is
renormalizable. The strong short-distance fluctuations of the
Proca potentials have been ``carried away'' by the discarded derivative term.

We have carried out this program in the easiest case:
  the coupling of a string-localized vector potential to a conserved 
  classical current \cite{MRS}. The interacting field is
  found to be  
$$A^{\rm int}_\mu(x,e)=A_\mu(x,e) + A^{\rm cl}_\mu(x) + \partial_\mu
\phi^{\rm cl}(x,e),$$
where $A_\mu(x,e)$ is the free quantum field, and $A^{\rm cl}_\mu(x)$ and
$\phi^{\rm cl}(x,e)$ are the classical retarded solutions associated with the
current $j_\mu(x)$ and with the source $j_e(x):= \int\nolimits_0^\infty
ds \, j_\mu(x+se)e^\mu$, respectively. Clearly, the interacting field
strength is $e$-independent.

See \sref{s:DVZWW} for the analogous case of 
massive and massless gravity. 

The same strategy to secure causality of the perturbation
  theory applies whenever the string-dependence of an
interaction term $L_{\rm int}(e)$ is a total derivative, so that
the classical action $S_{\rm int}(e)$ is
$e$-independent. In the case of massive 
tensor fields, one may have the form $L_{\rm int}(x,e)= L_{\rm
  int}^{\rm p}(x)+\partial_\mu V^\mu(x,e)$, where $L_{\rm int}^{\rm
  p}$ is point-localized but non-renormalizable; its UV-divergences
are absorbed by the derivative 
term such that $L_{\rm int}(e)$ is string-localized and renormalizable.
Interactions of massless particles do not possess an equivalent
point-local Lagrangian in the Hilbert space, but for the
$e$-independence of the action it is sufficient that
$\partial_{e^\kappa} L_{\rm int}(e) = \partial_\mu Q^\mu_\kappa$. 

However, the $e$-independence of the causal S
matrix requires at the quantum level that the time-ordering
can be defined in such a way that total derivatives are preserved. 
These conditions impose already in lowest orders certain constraints
on the possible interactions, that are all realized in the Standard
Model: the Lie algebra structure of 
cubic couplings of several species of vector bosons \cite{S16}; the
presence of a Higgs field when there are non-Abelian massive vector
bosons \cite{S16,MS}; and the chirality of their coupling to fermions
\cite{GMV}. In scalar massive QED, the cubic part of the string-local
minimal coupling induces also the quartic part \cite{S16},
so that the full fibre-bundle-like structure of the quantum
  theory turns out to be a consequence of imposing $e$-independence of
  the unitary S matrix, hence of positivity and causality, rather than
  a classical local gauge symmetry. We wonder whether this remarkable
  feature extends also to higher spin and gravitational couplings, possibly demanding
  additional couplings to lower spin fields.

These observations for $s=1$ resp.\ speculations for $s=2$
are very analogous to the analysis by Scharf et al.\ \cite{DSf,DS,Sf,DGSV} 
(pursued in the gauge-theoretic indefinite-metric and 
point-localized setting) where the fibre bundle structure
  and the presence of a Higgs boson are
  consequences of BRST invariance of causal perturbation theory with 
  self-interacting massive vector bosons. For spin 2,
  BRST invariance requires to supplement the cubic self-interaction of
  perturbative gravity in a unique way by higher-order terms that
  eventually sum up to the full Einstein gravity \cite{Sf}.

Indeed, like BRST invariance, also the condition of
$e$-independence can be formulated in a cohomological manner. Yet, the 
precise relation between gauge invariance and string-independence
remains to be explored. But beyond this analogy, it becomes clear that
the role of the Higgs boson is not the generation of the mass, but the
preservation of the renormalizability and locality under the
constraints imposed by positivity \cite{S16,MS}.  

\subsection{Properties of string-localized fields}

Let us turn back to \eref{AFe}, that we shall henceforth write in a
short-hand notation:
\bea\label{Ie}
A_\mu(x,e) = (I_e F_{\mu\nu})(x)e^\nu
\eea
where $I_e$ stands for the integration over the string in the direction $e$.
This potential is certainly not a fundamental field variable; it is some
``useful function'' of the field strength. Exploiting
the freedom to define different fields in terms of the same creation
and annihilation operators, we understand string-localized potentials 
mainly as a device to set up renormalizable interaction terms that are
equivalent to but better behaved than their
non-renormalizable point-localized counterparts.

It is also clear that string-localization is not a
feature of the associated particles, which are always the same massive
or massless particles specified by the Wigner representation. 
(The only exception would be particles in the infinite-spin 
representations \cite{MSY,LMR,R17}, whose fields are ``intrinsically
string-localized'', i.e., not representable as integrals like
\eref{AFe}. This case is beyond the scope of the present paper.)

Working with $A_\mu(e)$ is not in conflict
with the principle of causality, which is as imperative in
relativistic quantum field theory as Hilbert space positivity. While
their field strengths are point-localized, the string-localized
potentials satisfy causal commutation relations according to their
localization: two such operators commute whenever every point on 
the string $x+\RR^+_{0}e$ is spacelike to every point on the other string
$x'+\RR^+_{0}e'$. If the strings are chosen spacelike, such pairs of
spacelike separated strings 
are abundant. In this work, we assume
$e^2=-1$ without loss of generality.

\newpage

It follows from \eref{AFe} (and later generalizations
involving tensor fields and/or iterations of the integral operation
$I_e$) that the Poincar\'e transformations of string-localized fields are 
\bea\label{covariance} 
U(a,\Lambda) A_{\mu_1\dots\mu_r}(x,e) U(a,\Lambda)^* =
\Lambda^{\nu_1}{}_{\mu_1}\cdots \Lambda^{\nu_r}{}_{\mu_r}A_{\nu_1\dots\nu_r}(a+\Lambda x,\Lambda
e),
\eea
i.e., the direction of the string is transformed along with its apex
$x$ and the tensor components of the field tensor. Unlike a fixed
string direction, a transforming direction does not violate
covariance. See \rref{r:axial} for further comments, comparing with
``axial gauges'' with fixed directions. 

We shall write \tpf s throughout as
\bea\label{tpf-kernel}
(\Omega,X(x)Y(y)\Omega)= \int d\mu_m(p) \cdot 
e^{-ip(x-y)}\cdot \merw{m}{X,Y}(p),
\eea
where $d\mu_m(p)=\frac{d^4p}{(2\pi)^3}\delta(p^2-m^2)\theta(p^0)$.
Our sign convention is $\eta_{00}=+1$.

From the \tpf\ of the field strengths, one can compute the \tpf\ of
their string-localized potentials of any mass $m>0$ or $m=0$:\footnote{Our choice to consider correlations between fields with
  strings $-e$ and $e'$ is a convention to simplify notations, that
  will pay off when it comes to higher spin.}
\bea\label{AA1} \merw{m}{A_\mu(-e),A_\nu(e')} = -E(e,e')_{\mu\nu}(p),
\eea
where $E(e,e')_{\mu\nu}(p)$ is the distribution in $p$, $e$, and $e'$
\bea\label{def-E} E(e,e')_{\mu\nu}(p):= 
\eta_{\mu\nu} -\frac{p_\mu
  e_\nu}{(pe)_+}-\frac{e'_\mu p_\nu}{(pe')_+} + \frac{(ee')p_\mu
  p_\nu}{(pe)_+(pe')_+} .
\eea
The denominators $1/(pe)_+= 1/((pe + i0))$ arise from the integrations
$\int_0^\infty du\, e^{ip(x+ue)} = \frac i{p\cdot e +i0} e^{ipx}$.
\eref{AA1} should be compared with the massless \tpf\ in the
Gupta-Bleuler (Feynman gauge) approach:
\bea\label{AKAK}
\merw{0}{\AK_\mu,\AK_\nu}= -\eta_{\mu\nu}\,,
\eea 
the massless \tpf\ in the Coulomb gauge $A^{\rm C}_0=0$, $\vec\nabla
\vec A^{\rm C}=0$:
\bea\label{ACAC}
\merw{0}{A^{\rm C}_i,A^{\rm C}_j} = \delta_{ij}-\frac{p_ip_j}{\vert
\vec p\vert^2}\,,
\eea
and the \tpf\ of the Proca potential related to the massive field strength
by the Proca field equation $m^2\,\AP_\mu(x)=\partial^\nu F_{\mu\nu}(x)$:
\bea\label{APAP}\merw{m}{\AP_\mu,\AP_\nu} = - \eta_{\mu\nu} +
\frac{p_\mu p_\nu}{m^2} =: - \pi_{\mu\nu}(p).
\eea 

\eref{AKAK} is point-localized but obviously indefinite, \eref{ACAC} is
positive\footnote{By
  ``positive'', it is actually understood ''positive-semidefinite'',
  accounting for the null states due to equations of motion.} 
but non-covariant and completely non-local \cite{W}, \eref{APAP} is
point-localized and positive but has short-distance dimension
$d_{UV}=2$ due to the momenta in the numerator. Moreover, it obviously
does not admit a massless limit. 

In contrast, the string-localized \tpf\ \eref{AA1} is positive and
covariant (in the sense of \eref{covariance}) and it has
$d_{UV}=1$ for all $m\geq0$, and the massless case is smoothly connected
to the massive case. 

All \tpf s produce the same \tpf\ for the field strengths
$F_{\mu\nu}=\partial_\mu A_\nu-\partial_\nu A_\mu$:
\bea\label{FF1}
\merw{m}{F_{\mu\nu}, F_{\kap\lambda}} = 
-p_\mu p_\kap\eta_{\nu\lambda}+ p_\nu p_\kap\eta_{\mu\lambda}
+p_\mu p_\lambda\eta_{\nu\kap} -p_\nu p_\lambda\eta_{\mu\kap},
\eea
because \eref{AKAK}--\eref{AA1} all differ only by terms proportional
to $p$ that are ``killed'' by the curl. 

Indeed, the string-localized potential $A_\mu(x,e)$ for $e=(1,0,0,0)$
coincides with the Coulomb gauge potential $A^{\rm C}_\mu$.
The well-known non-locality of the Coulomb gauge potential \cite{W} 
reflects the fact that two timelike strings are never spacelike separated. 
It may be interesting to notice that one can average the 
potential $A_\mu(x,e)$ in $e$ over the spacelike sphere with $e^0=0$. The
resulting potential is again the Coulomb gauge potential
\eref{ACAC}. (Similar statements also hold for $s>1$, $m=0$, but the averaging
must be very carefully performed.)

\begin{remark}
\label{r:axial}
By definition, or by inspection of the \tpf, $A_\mu(x,e)$ satisfies
the relation $e^\mu A_\mu(x,e)=0$. This is the axial gauge condition
if $e$ is spacelike; indeed, \eref{AA1} coincides with the respective
gauged \tpf s. We emphasize that these gauge conditions are not used
to reduce the degrees of freedom before quantization, but that instead
the potentials for all $e$ coexist simultaneously on the Fock space of
the field strength, and they covariantly transform into each other
according to \eref{covariance}.  

By specifying the \tpf\ for spacelike $e$ as a distribution rather
than a function with a singularity, we reveal the manifestly
string-localized representation \eref{AFe} of the axial gauges, and
we discover the mutual commutativity of axial gauge potentials for
different directions.
\end{remark}

Unlike the Proca potential, the string-localized potential is not
conserved. Let
\bea a(x,e):= -m\inv\cdot \partial^\mu A_\mu(x,e).
\eea
One sees from the \tpf\ \eref{AA1} that in spite of the factor $m\inv$,
$a(x,e)$ is regular in the massless limit:
\bea\label{aA1}\merw{m}{a(-e), A_\nu(e')} &=& im
\Big(\frac{e_\nu}{(pe)_+}-
  \frac{(ee')p_\nu}{(pe)_+(pe')_+}\Big) ,\\
\label{aa1}\merw{m}{a(-e),a(e')} &=& 1 -m^2\frac{(ee')}{(pe)_+(pe')_+} .\eea
At $m=0$, $A_\mu(e)$ and $a(e)$ decouple, and \eref{aa1} is independent of
$e$ and $e'$, hence $\varphi(x)=a(x,e)\vert_{m=0}$ is independent of
$e$. Its one-particle state is the remnant of the massive particle
state with longitudinal angular momentum, decoupled from the massless
string-localized Maxwell potential. 

Finally, we have the identity (underlying \xref{x:QED}) 
\bea\label{expans1}\AP_\mu(x)= A_\mu(x,e) - m\inv \partial_\mu
a(x,e),\eea
that can also be seen from the definitions, using the Proca field
equations, or from the \tpf s.

\subsection{Spin $\geq 2$: DVZ discontinuity and Weinberg-Witten theorem}
\label{s:DVZWW}

The case of spin 2 (and higher) exhibits several new features as compared to spin
1, apart from the analogous problems with positivity, covariance,
short-distance dimension and massless limit. We here only sketch some
pertinent results for spin 2, that are proven for general integer spin
in \sref{s:string}. 

The first new feature is the DVZ observation 
due to van Dam and Veltman \cite{vDV} and Zakharov \cite{Z}, that in
interacting models with $s\geq 2$, scattering amplitudes are
discontinuous in the mass at $m=0$, i.e., the scattering on massless
gravitons (say) is significantly different from the scattering on
gravitons of a very small mass. The DVZ discontinuity has been used to
argue that, by measuring the deflection of light in a gravitational
field, gravitons must be exactly massless.  

The second new feature is the Weinberg-Witten theorem \cite{WW,Lp,L}
about the higher-spin massless case. It states that for $s\geq2$, 
no Lorentz covariant point-localized \set\ exists
such that the Poincar\'e generators are moments of its zero-components:
\bea\label{generators} P_\sig \stackrel!= \int_{x^0=t} d^3\vec x \,
T_{0\sig}, 
\qquad
M_{\sig\tau}  \stackrel!= \int_{x^0=t} d^3\vec x \, (x_\sig T_{0\tau} - x_\tau
T_{0\sig}).
\eea
The absence of a Lorentz covariant \set\ also obstructs the
semiclassical coupling of massless higher spin matter to
gravity. More precisely, because there exists a \set\ in
  terms of potentials whose Lorentz transforms involve gauge 
  transformations, the problem is shifted to the challenge of finding gauge
  invariant couplings to the gravitational field.\footnote{We thank
    the referee for this more precise formulation of the issue.} 

The DVZ discontinuity can be traced back to the fact that for spin 2
(or higher), the massless limit of the massive field strength
\bea\label{F2}
F_{[\mu\nu][\kap\lam]}=\partial_\mu\partial_\kap A_{\nu\lam}-\partial_\nu\partial_\kap
A_{\mu\lam}-\partial_\mu\partial_\lam
A_{\nu\kap}+\partial_\nu\partial_\lam A_{\mu\kap}
\eea
exists but differs from the massless field strength. This is seen explicitly
by inspection of the \tpf s, which in the massive case is the curl
(taken in all indices) of the positive \tpf\ of the spin-2 Proca field\footnote{Albeit historically incorrect, we
  adopt the name ``Proca field'' also for higher spin.}
\bea\label{APAP2}
\merw{m}{\AP_{\mu\nu}, \AP_{\kap\lam}}
=\frac12\Big[\pi_{\mu\kap}\pi_{\nu\lam}+\pi_{\mu\lam}\pi_{\nu\kap}
\Big]
-\frac13\pi_{\mu\nu}\pi_{\kap\lam},
\eea 
and in the massless case is the curl of 
\bea\label{AKAK2}
\merw{0}{\AK_{\mu\nu}, \AK_{\kap\lam}}
=\frac12\Big[\eta_{\mu\kap}\eta_{\nu\lam}+\eta_{\mu\lam}\eta_{\nu\kap}
\Big]
-\frac12\eta_{\mu\nu}\eta_{\kap\lam},
\eea 
where $\AK$ is the Feynman gauge potential. We emphasize, however, 
that the massless field strength is autonomously defined on the
positive Fock space over the helicity $h=\pm 2$ Wigner
representations, and the indefinite Feynman gauge potential does not
exist on this Hilbert space. 

Applying the curls to both \eref{APAP2} and \eref{AKAK2}, the
difference between the tensors $\pi_{\mu\nu}$ (\eref{APAP}) and $\eta_{\mu\nu}$ disappears; but the
different coefficient $-\frac13$ vs.\ $-\frac 12$ of the last term
survive. These coefficients are intrinsic features of the underlying
massive and massive representations: they fix the number of linearly
independent states in the one-particle spaces whose scalar product is
given by the \tpf s of the field strengths ($2s+1=5$ in the massive
case, $2$ in the massless case). 

In order to analyze the DVZ discontinuity with the help of
string-localized fields, we have to properly decouple in the massless
limit the lower helicities $h=0,\pm1$, that all contribute to the
massive field, from the helicities $h=\pm2$ of the graviton. Let us
first study the decoupling. 

The massive string-localized potential is  
\bea\label{a2-def}A_{\mu\nu}(x,e):=\big(I_e^2
F_{[\mu\kap][\nu\lam]}\big)(x)e^{\kap} e^{\lam}\,
\eea
such that its double curl gives back the massive field strength. Its \tpf\ is
\bea\label{aa2-sl} 
\merw{m}{A_{\mu\nu}(-e),A_{\kap\lam}(e')}=\frac12\Big[E(e,e')_{\mu\kap}E(e,e')_{\nu\lam}+(\kap\lra\lam)\Big] - \frac13
E(e,e)_{\mu\nu}E(e',e')_{\kap\lam}.\quad
\eea

Unlike the spin-2 Proca potential, the string-localized potential is not
conserved. We define the escort fields
\bea\notag a^{(1)}_\mu(x,e)&:=& -m\inv\partial^\nu
A_{\mu\nu}(x,e), \\
\label{esc2-def}a^{(0)}(x,e)&:=& -m\inv\partial^\mu a^{(1)}_\mu(x,e).
\eea
They are also regular at $m=0$ because $\partial^\nu
F_{[\mu\kap][\nu\lam]} =-m^2 \FP_{[\mu\kap]\lam}$ (the partial field
strength \cite{Fz}), and $\partial^\mu \FP_{[\mu\kap]\lam}= -m^2
\AP_{\kap\lam}$. Moreover, the identity 
\bea\label{TrA}
a^{(0)}(x,e) = -\eta^{\mu\nu} A_{\mu\nu}(x,e)
\eea
holds, as well as the decomposition
\bea\label{AAP}
\AP_{\mu\nu}(x)=A_{\mu\nu}(x,e) -m\inv\big(\partial_\mu
a^{(1)}_\nu(x,e)+\partial_\nu
a^{(1)}_\mu(x,e)\big)+m^{-2}\partial_\mu\partial_\nu a^{(0)}(x,e).
\eea

From \eref{aa2-sl} all other \tpf s can be computed by descending with
\eref{esc2-def}. One finds that $a^{(1)}_\mu(e)$ 
decouples from $A_{\mu\nu}(e)$ and from
$a^{(0)}(e)$ in the limit $m\to0$, but the latter two do not decouple
from each other: 
\bea\label{Aa20}
\merw{0}{a^{(0)}(-e),A_{\mu\nu}(e')} &=& 
- \frac13 E(e',e')_{\mu\nu}(p) \\
\label{a20a20}\merw{0}{a^{(0)}(-e),a^{(0)}(e')}&=&\frac23.
\eea 

In order to decouple them, let
\bea\label{A22-mom}
A^{(2)}_{\mu\nu}(x,e) := A_{\mu\nu}(x,e) +\frac12
E_{\mu\nu}(e,e) a^{(0)}(x,e),\eea
where the integro-differential operator  
\bea\label{idiff}
E(e,e)_{\mu\nu} = \eta_{\mu\nu}+ 
\big(e_\nu\partial_\mu+e_\mu\partial_\nu\big) I_e 
+e^2 \partial_\mu\partial_\nu I_e^2
\eea
acts by multiplication with $E(e,e)_{\mu\nu}(p)$ and $E(e,e)_{\mu\nu}(-p) =
E(-e,-e)_{\mu\nu}(p)$ on the creation and annihilation parts,
respectively. With this redefinition, the decoupling is exact at
$m=0$, and at all $m\geq0$
\bea\label{AA2-sl} 
\merw{m}{A^{(2)}_{\mu\nu}(-e),A^{(2)}_{\kap\lam}(e')}=\frac12\Big[E(e,e')_{\mu\kap}E(e,e')_{\nu\lam}+(\kap\lra\lam)\Big] - \frac12
E(e,e)_{\mu\nu}E(e',e')_{\kap\lam},\quad
\eea
which is exactly \eref{aa2-sl} except for the 
the coefficient $-\frac12$ of the last term. Thus, at $m=0$, 
$A^{(2)}$ coincides with the string-localized potential 
\bea\label{A2F2}A^{(2)}_{\mu\nu}(x,e)=\big(I_e^2
F^{(m=0)}_{[\mu\kap][\nu\lam]}\big)(x)e^{\kap} e^{\lam}
\eea
associated with the massless field strength. The appropriately
normalized field $A^{(1)}_\mu(e):=\sqrt2\cdot a^{(1)}_\mu(e)$ converges
at $m=0$ to the string-localized Maxwell potential, and
$A^{(0)}(e):=\sqrt{3/2}\cdot a^{(0)}(e)$ converges to the $e$-independent
scalar field $\varphi(x)$. Thus, the fields $A^{(r)}(e)$ with $r=0,1,2$ parametrize
the exact decoupling of helicities $h=\pm r$ at $m=0$.

The result generalizes to arbitrary integer spin: Linear combinations
of $A_{\mu_1\dots\mu_s}(e)$ and its lower rank escort fields acted
on by the operator $E(e,e)_{\mu\nu}$ yield fields $A^{(r)}_{\mu_1\dots\mu_r}(e)$
for every $0\leq r\leq s$ which in the massless limit are decoupled
potentials of helicity $h=\pm r$ field strengths. $A^{(0)}(e)$ becomes
independent of $e$ and is the massless scalar field. Of course, the total
number $2s+1$ of one-particle states is preserved. All these
potentials have short-distance dimension $d_{UV}=1$ at $m\geq0$ and
are suited for setting up renormalizable perturbation theory. 

Now, returning to the DVZ problem, we may couple perturbative massive
gravity in a Minkowski background to a conserved stress-energy source by  
\bea\label{Se}
S_{\rm int}(e) = \int d^4x\, A_{\mu\nu}(x,e)T^{\mu\nu}(x).
\eea
Because by \eref{AAP}, $A_{\mu\nu}(e)$ differs from $\AP_{\mu\nu}$ only
by derivatives, the action is independent of $e$. At $m>0$, all five
states of the graviton couple to the source. In the limit $m\to0$, we
have by \eref{A22-mom} 
$$A_{\mu\nu}(x,e) = A^{(2)}_{\mu\nu}(x,e) -
\sqrt{1/6}\,\eta_{\mu\nu} \,\varphi(x) + \hbox{derivatives},$$
where $\varphi(x)=\sqrt{3/2}\,\lim_{m\to0} a^{(0)}(x,e)$ is the
string-independent massless
scalar field decoupled from the helicity-2 potential $A^{(2)}(x,e)$. Thus, 
\bea\label{Se0}
\lim_{m\to0} S_{\rm int}(e) = \int d^4x\,
A^{(2)}_{\mu\nu}(x,e)T^{\mu\nu}(x) - \sqrt{1/6} \int d^4x\,
\varphi(x)\, T_\mu^\mu(x).
\eea
The first (pure massless gravity) contribution is independent of
$e$ by virtue of \cref{c:m=0}.

We have thus explicitly identified the scalar field that is responsible for
the DVZ discontinuity, as the limit of the escort field on the massive
Hilbert space. This is formally equivalent with Zakharov's reading who
writes instead the massless coupling
$\AK_{\mu\nu}(x,e)T^{\mu\nu}(x)$ as the limit of the massive coupling
plus a compensating scalar ghost \cite{Z}; but we emphasize that our
reading does not involve unphysical ghost degrees of freedom. 

The same decoupling of helicities in terms of $A^{(r)}(e)$ with
$0\leq r\leq s$ also allows to construct a massless stress-energy tensor for
arbitrary helicity. It is quadratic in $A^{(r)}(e)$, hence also
string-localized. There is no conflict with the Weinberg-Witten
theorem, that assumes point-localized fields. 

\newpage

It was already discussed in \cite{Lp} that the Weinberg-Witten theorem
does not exclude non-local densities. The string-localized \set s
realize this possibility. As compared to other proposals
\cite{FV,V1,V2,L,BBS} evoking an interplay of infinitely many spins,
M-theory, and non-commutative geometry, the  
string-localized \set s of \pref{p:set} for every pair of helicities 
$h =\pm s$ are perhaps the most conservative way around the Weinberg-Witten
theorem. They are even ``less non-local'' than the examples with unpaired
helicities proposed in \cite{Lp}. 

We are presently investigating how
they may be used to (semiclassically) couple massless higher spin matter to
gravity. Similar as in \xref{x:QED} or \eref{Se}, this
  requires to identify string-independent actions involving a
  string-dependent \set. A straightforward ansatz
  $h_{\mu\nu}T^{\mu\nu}(e)$ with $h$ a point-localized massive or
  string-localized massless tensor may require additional terms
  to ensure string-independence. Cf.\ also \sref{s:overview}.

The construction of the string-localized massless \set\ proceeds along
the following lines. For more details, see \sref{s:string}. 

For $s=2$ one may start at $m>0$ from the Pauli-Fierz Lagrangian
\cite{FP} whose field equations are the spin-2 Proca (or rather Fierz)
equations \cite{Fz}. The Hilbert \set\ is defined by the variation
w.r.t.\ the metric of its generally covariant version. One may as well
start from a simpler ``reduced'' massive \set\ whose densities differ
by spatial derivatives; it therefore yields the same Poincar\'e 
generators \eref{generators}, and is as good for the purpose. The
reduced \set\ easily generalizes to arbitrary spin, in fact without
the need for a free higher spin Lagrangian. Since the \tpf\ determines
the commutator, one can (and must) verify that the generators
implement the correct infinitesimal Poincar\'e transformations. 

(To our surprise, the generators associated with 
the \set\ given by Fierz \cite{Fz} implement the correct
translations, but wrong Lorentz transformations.)

By inserting the decomposition 
\bea\label{expans2}\AP_{\mu\nu}(x) = A_{\mu\nu}(x,e) - m\inv\big(\partial_\mu a^{(1)}_\nu
+\partial_\nu a^{(1)}_\mu\big)(x,e) + m^{-2} \partial_\mu\partial_\nu a^{(0)}(x,e),\eea
respectively its generalization \eref{exps} to arbitrary spin, into the
reduced \set, it turns out that all contributions with negative 
powers of the mass multiplied by derivatives of escort fields can be
combined into spatial derivative terms that do not contribute to the
generators, see \aref{a:lemma}. Discarding these terms, one arrives at
a ``regular'' \set\ that admits a massless limit. It is quadratic in
escort fields $a^{(r)}_{\mu_1\dots\mu_r}$ for all $0\leq r\leq s$; but
not yet of much use because these fields are still coupled through
their traces at $m=0$.  

By expressing the massless escort fields $a^{(r)}_{\mu_1\dots\mu_r}$ in terms
of the decoupling massless potentials $A^{(r)}_{\mu_1\dots\mu_r}$, one
may again discard contributions from the regular \set\ that do not
contribute to the generators at $m=0$. The resulting \set\ is
quadratic in the massless potentials
$A^{(r)}_{\mu_1\dots\mu_r}$. Because the latter mutually commute, this
\set\ is a direct sum of massless \set s for all $0\leq r\leq
s$. These separately yield by \eref{generators} the generators of the helicity
$h=\pm r$ subrepresentations of the massless limit of the massive
spin $s$ representation.

\section{General $s$: Preliminaries on point-localized fields}
\label{s:point}
\setcounter{equation}{0}

\subsection{Massive case}
The massive Proca field $\AP_{\mu_1\dots\mu_s}$ of spin $s$ is a completely
symmetric traceless and conserved tensor field satisfying the
Klein-Gordon equation: 
\bea\label{eom}
\eta^{\mu_i\mu_j}\AP_{\mu_1\dots\mu_s}=0,\qquad \partial^{\mu_j}
\AP_{\mu_1\dots\mu_s}=0,\qquad (\square +m^2)\AP_{\mu_1\dots\mu_s}=0,
\eea
with \tpf\ 
\bea\label{tpf} \merw{m}{\AP_{\mu_1\dots\mu_s},\AP_{\nu_1\dots\nu_s}} =
(-1)^s\sumno_{2n\leq s} \wt {\beta^s_n}\,
(\pi_{\mu\mu})^n(\pi_{\nu\nu})^n (\pi_{\mu\nu})^{s-2n}. 
\eea
The sum extends over all inequivalent attributions of the
available indices of the given form, namely either $\pi_{\mu_i\mu_j}$,
$\pi_{\nu_i\nu_j}$, or $\pi_{\mu_i\nu_j}$, to the schematically displayed factors.
Due to the factors of $p$ in the numerator, the short-distance
  dimension of $\AP$ is $d_{UV}=s+1$.

The coefficients $\wt {\beta^s_n}=\frac{n!}{s!(\frac12-s)_n}$ ensure the
vanishing of the traces, while the 
fact that $\pi_{\mu\nu}(p)p^\nu=0$ 
ensures the vanishing of the divergences. (The alternating overall sign is
  due to our sign convention of the metric.)

The same formula can also be
derived from the $(m,s)$ Wigner representation:
\bea\label{APu} 
\AP_{\mu_1\dots\mu_s} = \int d\mu_m(p)
\Big[e^{ipx}\sum\nolimits_{a=1}^{2s+1}
  u^a_{\mu_1\dots\mu_s}(p)a_a^*(p) + \hbox{h.c.}\Big]
\eea
where the tensors $u^a_{\mu_1\dots\mu_s}(p)$ intertwine the $(m,s)$ Wigner
representation of the Lorentz group with the symmetric traceless
tensor representation \cite{W}, and the spin indices $a$ are
summed over. The coefficients $\wt {\beta^s_n}$ in \eref{tpf} arise due to
the projection operator (involved in the intertwiners
$u^a_{\mu_1\dots\mu_s}$) onto the spin $s$ 
representation in the $s$-fold tensor product of vector
representations of the little group $\SO(3)$ (= traceless symmetric
tensors in $(\CC^3)^{\otimes s}$ \cite[Eq.~(1.13)]{Gr}). 

To keep track of the combinatorics for general $s$, it will be 
advantageous to trade the indices for an ``orientation vector''
$f\in\RR^4$ and write
$$X(f)\equiv X_{\mu_1\dots\mu_r}f^{\mu_1}\dots f^{\mu_r}, $$
when $X$ is a symmetric rank $r$ tensor. Then the divergence
$(\partial X)_{\mu_2\dots\mu_r}
:= \partial^{\mu_1}X_{\mu_1\dots\mu_r}$ and the trace 
$(\Tr X)_{\mu_3\dots\mu_r}:= \eta^{\mu_1\mu_2} X_{\mu_1\dots\mu_r}$ are
given by 
$$r\cdot(\partial X)(f)= (\partial_x\cdot\partial_f X(f)),\qquad r(r-1)
\cdot(\Tr
X)(f) = \square_f X(f).$$
In this notation, the Proca \tpf\ \eref{tpf} is written
\bea
\merw{m}{\AP(f),\AP(f')} = (-1)^s\sumno_{2n\leq s} \beta^s_n\, (f^t\pi f)^n
(f'^t\pi f')^n (f^t\pi f')^{s-2n}
\eea
whose coefficients differ from $\wt
{\beta^s_n}$ by a counting factor of equivalent terms:
\bea\label{beta}
\beta^s_n= \Big[{\binom s{2n}} (2n-1)!!\Big]^2 (s-2n)! \cdot \wt {\beta^s_n} =
\frac{1}{4^nn!}\frac{s!}{(s-2n)!}\frac{1}{(\frac12-s)_n}. 
\eea
In $D$ dimensions (where $\Tr(\pi) = D-1$, little group $\SO(D-1)$),
the Pochhammer symbol $(\frac12-s)_n$, arising from the projection
onto traceless symmetric tensors in $(\CC^{D-3})^{\otimes s}$, would
be replaced by $(\frac{5-D}2-s)_n$.

\subsection{Massless case}
\label{s:m=0}
For the massless case, point-localized covariant potentials with a
positive \tpf\ do not exist. From the pair of Wigner representations
$(m=0,h=\pm s)$ one can construct point-localized covariant field strengths  
$F^{(s)}_{[\mu_1\nu_1]\dots[\mu_s\nu_s]}$
whose \tpf\ are the curls of the indefinite \tpf\ 
\bea\label{tpfK} \merw{0}{\AK_{\mu_1\dots\mu_s},\AK_{\nu_1\dots\nu_s}} =
(-1)^s\sumno_{2n\leq s} \wt {\gamma^s_n}\,
(\eta_{\mu\mu})^n(\eta_{\nu\nu})^n (\eta_{\mu\nu})^{s-2n}, 
\eea
(notation as in \eref{tpf}) or equivalently,
\bea
\merw{0}{\AK(f),\AK(f')} = (-1)^s\sumno_{2n\leq s} \gamma^s_n\, (f^t\eta f)^n
(f'^t\eta f')^n (f^t\eta f')^{s-2n}.
\eea
The coefficients are
\bea\label{gamma}
\gamma^s_n= \Big[{\binom s{2n}} (2n-1)!!\Big]^2 (s-2n)! \cdot \wt {\gamma^s_n} =
\frac{1}{4^nn!}\frac{s!}{(s-2n)!}\frac{1}{(1-s)_n}. 
\eea
In $D$ dimensions (little group $\Eu(D-2)=\SO(D-2)\ltimes\RR^{D-2}$
with $\RR^{D-2}$ represented trivially), the Pochhammer symbol
$(1-s)_n$, arising from the projection onto traceless symmetric
tensors in $(\CC^{D-2})^{\otimes s}$, would
be $(\frac{6-D}2-s)_n$. 

\section{String-localized fields: general integer spin $s$}
\label{s:string}
\setcounter{equation}{0}

Throughout this section the spin $s$ is fixed, and does not always appear
explicitly in the notation; i.e., fields like $a^{(r)}$ or numbers like
$\beta^{rr'}_{nn'}$ will depend also on $s$.

Let $\AP_{\mu_1\dots\mu_s}$ and $F_{[\mu_1\nu_1]\dots[\mu_s\nu_s]}$
be the Proca potential and its field strength. Let $e$ be a (spacelike)
unit vector. We introduce the symmetric string-localized potential  
\bea\label{as}
a^{(s)}_{\mu_1\dots\mu_s}(x,e):=
(I_e^s F_{[\mu_1\nu_1]\dots[\mu_s\nu_s]})(x)e^{\nu_1}\dots e^{\nu_s}
\eea
defined on the Wigner Fock space over the $(m,s)$ Wigner
representation. It is another potential for
$F_{[\mu_1\nu_1]\dots[\mu_s\nu_s]}$, but unlike
$\AP_{\mu_1\dots\mu_s}$, it is neither traceless nor conserved.

We display the \tpf\ $\merw{m}{a^{(s)}(-e),a^{(s)}(e')}$. Every factor
$\pi$ in \eref{tpf} is hit by two of the matrices $J(p,e')$ or
$\overline{J(p,-e)}= J(p,e)$. We therefore define
\bea\label{EJJ}
E_{\mu\nu}(e_1,e_2)(p) :=
(J(p,e_1)\pi(p)J(p,e_2)^t)_{\mu\nu}=(J(p,e_1)\eta J(p,e_2)^t)_{\mu\nu}
\eea
which is precisely the distribution defined in \eref{def-E},
and abbreviate (for $f,f'\in \RR^4$)
$$E_{ff}\equiv f^t E(e,e)f, \quad
E_{ff'}\equiv f^t E(e,e')f',\quad E_{f'f'}\equiv f'^t E(e',e')f'.$$
Then we have from \eref{tpf}:
\bea\label{ass}
\merw{m}{a^{(s)}(-e)(f),a^{(s)}(e')(f')}
= (-1)^s\sumno_n \beta^s_n \, (E_{ff})^n(E_{f'f'})^n(E_{ff'})^{s-2n}.
\eea

Because $E(e,e')(p)$ is a homogeneous function of $p$, the
  short-distance dimension of $a^{(s)}$ is $d_{UV}=1$.

\subsection{Escort fields}
In order to establish the relation between $\AP$ and $a^{(s)}$, and to
control the massless limit, we introduce the escort fields for $0\leq r<s$
\bea\label{ar}
a^{(r)}_{\mu_1\dots\mu_r}(x,e):= -m\inv\cdot \partial^{\mu}
a^{(r+1)}_{\mu_1\dots\mu_r\mu}(x,e).
\eea

\begin{remark}
$a^{(s)}_{\mu_1\dots\mu_s}$ coincides with
the string-localized field denoted $A_{\mu_1\dots\mu_s}$ in
\cite{S15,S16,MO}. 
$a^{(r)}$ ($r<s$) are related to the escort fields $\phi^{(r)}$ introduced
there by derivatives of lower $\phi^{(q)}$ ($q<r$)
and an overall power of the mass: 
$$a^{(r)}_{\mu_1\dots\mu_s}(x,e) = m^{s-r} \sumno_{q\leq r}\partial_\mu\dots\partial_\mu \phi^{(q)}_{\mu\dots\mu}(x,e)
$$
where for each $q\leq r$ the sum extends over all $\binom rq$ inequivalent permutations
of the indices.
This can be seen from \cite[Eq.~(4)]{MO} by taking
divergences and using the Klein-Gordon equation. \lref{l:reg} below justifies our departure from the previous definition. 
\end{remark}

The definition \eref{as} involves the operations curl, contraction
with $e$ and string integration on each Lorentz index of
$\AP_{\mu_1\dots\mu_s}$. This means that $a^{(s)}_{\mu_1\dots\mu_s}$ arises
from \eref{APu} by multiplication of the intertwiner
$u^a_{\nu_1\dots\nu_s}$ with the matrix
\bea\label{matrix}
J_\mu{}^\nu(p,e)= \delta_\mu^{\nu} - \frac{p_\mu e^\nu}{(pe)_+},
\eea
in each Lorentz index. It is obviously
\bea\label{eJ=Jp=0}
J_\mu{}^{\nu}(p,e)p_\nu=0,\qquad e^\mu
J_\mu{}^{\nu}(p,e)=0.
\eea

\begin{coro}\label{c:ea=0}
The ``axial gauge'' condition (cf.\ \rref{r:axial})
$e^{\mu}a^{(r)}_{\mu\mu_2\dots\mu_r}(e)=0$ holds. 
\end{coro}
Proof: Evident from the second of \eref{eJ=Jp=0} and the definition
\eref{ar}. 
\qed

The following property secures the string-independence of
actions $S_{\rm int}(e)$ like \eref{Se1} and \eref{Se}.
\begin{coro}\label{c:e-dep}
The string-dependence of the string-localized potential
$a^{(s)}_{\mu_1\dots\mu_r}(e)$ is a sum of
derivatives: 
\bea\label{e-dep}
\partial_{e^\kappa} a^{(s)}_{\mu_1\dots\mu_s}(e) = 
\sumno_i\partial_{\mu_i} (I_e a^{(s)}_{\dots\mu_{i-1}\kappa\mu_{i+1}\dots}(e)). 
\eea
\end{coro}
Proof: Evident by computing the derivative $\partial_{e^\kappa}
J_{\mu}{}^\nu = -\frac{p_\mu}{(pe)_+}J_\kappa{}^\nu.$ 
\qed

This formula together with \eref{covariance} also explains why
  Lorentz transformations of axial gauge potentials at {\em fixed} $e$
formally involve an ``operator-valued gauge transformation''.

The conservation of $\AP$ means that $p^{\mu_i}
u^a_{\mu_1\dots\mu_s}(p)=0$ in \eref{APu}. Because $ip^\mu J_\mu{}^\nu
= ip^\nu - im^2\frac{e^\nu}{(pe)_+}$, it follows for $r\leq s$
\bea\label{aru}
a^{(r)}_{\mu_1\dots\mu_r}(x,e) = \int \!\!d\mu_m(p)
\Big[e^{ipx} \prod_{k=1}^r
J_{\mu_k}^{\,\,\nu_k}(p,e)\prod_{k=r+1}^s\frac{ime^{\nu_k}}{(pe)_+}
\,\sum_a u^a_{\nu_1\dots\nu_s}(p)a_a^*(p) + \hbox{h.c.}\Big]
\eea

\begin{lemma}
\label{l:reg}
The fields $a^{(r)}_{\mu_1\dots\mu_r}(x,e)$ ($0\leq r\leq s$) are regular in
the limit $m\to 0$. 
\end{lemma}
Proof: The \tpf s of $a^{(r)}_{\mu_1\dots\mu_r}$ arise from \eref{tpf}
by multiplying with the matrices $J$ and contracting with $im\frac e{(pe)_+}$
  according to \eref{aru}. By the first of \eref{eJ=Jp=0}, every
  matrix $J$ kills one singular factor $p_\mu p_\nu/m^2$ of
  \eref{tpf}, and the powers of $m$ coming with the contractions
  balance the remaining singularity. 
\qed

We have the (preliminary) decomposition of the point-localized field
$\AP_{\mu_1\dots\mu_s}$ into $a^{(s)}$ and its escort fields:

\begin{propo}\label{p:exps} The massive point-localized potential of spin
  $s$ can be written as
\bea\label{APA}
\AP_{\mu_1\dots\mu_s}(x) = \prodno_{k=1}^s
\big(\delta_{\mu_k}^{\nu_k}+m^{-2}\partial_{\mu_k}\partial^{\nu_k}\big) a^{(s)}_{\nu_1\dots\nu_s}(x,e).
 \eea
It decomposes into regular string-localized escort fields with inverse mass
  coefficients 
\bea\label{exps}
\AP_{\mu_1\dots\mu_s}(x) = a^{(s)}_{\mu_1\dots\mu_s}(x,e) +\sumno_{r< s}
(-m\inv)^{s-r}\partial_\mu\dots\partial_\mu a^{(r)}_{\mu\dots\mu}(x,e)
\eea
where for each $r<s$ the sum extends over all $\binom sr$ inequivalent permutations
of the indices. From this it is manifest that $\AP$ and $a^{(s)}$
  have the same field strength.
\end{propo}

Proof: In momentum space, the differential operator in \eref{APA}
is $\pi^{\otimes s}$. The identity follows from \eref{aru} (with
$r=s$), because $\pi J = \pi$ and $\pi^{\otimes s}
u^a=u^a$ since $\AP$ is conserved. 
The derivatives $m^{-2}\partial_\mu\partial^\nu$ involved in \eref{APA} turn
$a^{(r)}$ into $-m\inv \partial_\mu a^{(r-1)}$ by \eref{ar}. This gives \eref{exps}.
\qed

The string-localized fields $a^{(r)}$ are dynamically coupled among
each other. We have

\begin{propo}\label{p:eoms} The regular escort fields
  $a^{(r)}_{\mu_1\dots\mu_r}$ are coupled through the field equations
\bea\label{eoms} 
\partial^{\mu_1} a^{(r)}_{\mu_1\dots\mu_r} = -m\,
a^{(r-1)}_{\mu_2\dots\mu_r},\qquad
\eta^{\mu_1\mu_2}a^{(r)}_{\mu_1\dots\mu_r} = - a^{(r-2)}_{\mu_3\dots\mu_r}.
\eea 
\end{propo}
By the first equation, every
escort $a^{(r)}$ still ``contains'' all the lower escorts $a^{(r')}$
($r'<r$). The divergence will decouple in the massless limit from the
lower escorts, while the trace doesn't. Subtracting the traces would
instead bring back the coupling through the divergences. This is the
reason why the decomposition in \pref{p:exps} is only preliminary. 

Proof: The first equation is just the definition \eref{ar}. The second
follows from  
$$(J^t\eta J)^{\nu_1\nu_2} = \eta^{\nu_1\nu_2}
-\frac{p^{\nu_1}e^{\nu_2}+e^{\nu_1}p^{\nu_2}}{(pe)_+} +m^2 \frac{e^{\nu_1}e^{\nu_2}}{(pe)_+^2}$$
together with the fact that $\AP$ is traceless and conserved, hence
$p^\nu$ and $\eta^{\nu_1\nu_2}$ act trivially in \eref{aru}.  
\qed

\subsection{Decoupling in the massless limit}
\label{s:decoup}
The massless results of this section are equivalent to results obtained
recently by Plaschke and Yngvason \cite[Sect.~4A]{PY}. While these authors
consider Wigner intertwiners directly at $m=0$, we exhibit
smooth families of fields $A^{(r)}\vert_{m\geq 0}$. 

We turn to the task of a complete decoupling at $m=0$. We do this
by a study of the \tpf s. In a positive metric, decoupling
the \tpf s implies the decoupling of the field equations.

The \tpf s of the massive escort fields $a^{(r)}$ do not decouple. In
order to compute them efficiently, 
we cast \eref{aru} into the form of a ``generating functional'':
$$\sumno_{r\leq s}\binom sr
a^{(r)}(e)(f) = Z(f,e) := \AP(J_e^tf+m eI_e). $$
Here $I_e$ is the string integration, understood in this formula as an
operation acting on the field, and $J_e$ acts by multiplication with
$J_\mu{}^\nu(p,e)$ and its complex conjugate on the creation resp.\
annihilation part of the field.  
Then  
$$\merw{m}{Z(f,-e),Z(f',e')} = \sumno_{r,r'} \binom sr\binom s{r'}
\cdot \merw{m} {a^{(r)}(-e)(f),a^{(r')}(e')(f')}.$$
Given the l.h.s.\ as a function of $f$ and $f'$, the correlations
between $a^{(r)}$ and $a^{(r')}$ can be read off by
selecting the terms of the appropriate homogeneities in $f$ and in
$f'$. 

In order to compute the l.h.s., we have to contract each factor $\pi_{\mu\mu}$  
in \eref{tpf} twice with $(J(p,e)^tf-im e/(pe)_+)^\mu$, each factor 
$\pi_{\nu\nu}$ twice with $(J(p,e')^tf+im e'/(pe')_+)^\nu$, and each
factor $\pi_{\mu\nu}$ with both vectors. Because of the first of
\eref{eJ=Jp=0} and \eref{EJJ}, 
and because $(m e/(pe)_+)^t\pi(m e/(pe)_+) = -1 + O(m^2)$, all these
contractions are of the form $E+1 + O(m)$ resp.\ $E-1+O(m)$, and one
arrives at    
$$\merw{m}{Z(f,-e),Z(f',e')} = (-1)^s\sumno_{2n\leq s} \beta^s_n\, 
(E_{ff}+1)^n(E_{f'f'}+1)^n(E_{ff'}-1)^{s-2n} + O(m).$$
We get the massless \tpf s
\begin{propo}\label{p:arr'} At $m=0$, one has 
\bea\label{arr'}
\merw{0}{a^{(r)}(-e)(f),a^{(r')}(e')(f')}
= (-1)^r \sumno_{r-2n=r'-2n'} \beta^{rr'}_{nn'} (E_{ff})^n(E_{f'f'})^{n'}(E_{ff'})^{r-2n}
\quad
\eea
with 
\bea\label{brr'} \binom sr \binom s{r'}\cdot
\beta^{rr'}_{nn'} = \sumno_{m} \binom mn \binom m{n'}
\binom{s-2m}{r-2n}\cdot \beta^s_m.
\eea
In particular, $\merw{0}{a^{(r)}(-e),a^{(r')}(e')}=0$ if $r-r'$ is odd.
\end{propo}
Proof: \eref{brr'} are the coefficients of the respective terms of
homogeneity $r$ in $f$ and $r'$ in $f'$. \qed

One could also have computed \eref{arr'} by descending from
\eref{ass} with \eref{eoms} at $m>0$, and then taking $m\to 0$. 

We now set out to ``diagonalize'' the mixed \tpf s \eref{arr'} with
the help of the operator $E(e,e)_{\mu\nu}$ given in
\eref{idiff}. We write $E_{ff}\equiv f^t E(e,e)f$.

\begin{propo}\label{traceless}
The combinations
\bea\label{Ar}
A^{(r)}(f) = \sumno_{2k\leq r}\alpha^r_k \cdot (-E_{ff})^k a^{(r-2k)}(f)
\eea
are traceless at $m=0$ if and only if $\alpha^r_k = \alpha_r\cdot
\gamma^r_k$, with $\gamma^r_k$ given in \eref{gamma}. Only the
coefficient $\alpha_r$ (that will be used later for normalization) may
depend on $s$.  
\end{propo}
Proof: By applying $\square_f$ to \eref{Ar} and
noticing that $\Tr(E)=2+O(m^2)$, $\Tr(a^{(r)})=-a^{(r-2)}$, and
$E_{\nu}{}^{\mu}a^{(r)}_{\mu\mu_2\dots\mu_r}=
a^{(r)}_{\nu\mu_2\dots\mu_r}+O(m)$ because $\partial a^{(r)}=O(m)$ and
$ea^{(r)}=0$ (\eref{eoms} and \cref{c:ea=0}), one obtains the
recursion
$$\alpha^r_k = -
\frac{(r-2k+2)(r-2k+1)}{4k(r-k)}\alpha^r_{k-1}.$$
This is solved by $\frac{\alpha^r_k}{\alpha^r_0} = \gamma^r_k$. \qed 

Because the definition \eref{Ar} is upper triangular in $r$, the
inverse formula is of the same form. We did, however, not succeed to compute its
coefficients in closed form.

The operators $E_{ff}$ and $E_{f'f'}$ involved in the field
definitions produce the factors denoted with the same symbols (cf.\
\eref{ass}) in the \tpf s. Therefore, the correlations among
$A^{(r)}(f)\vert_{m=0}$ are of the same general form as \eref{arr'}
with different coefficients. Because $A^{(r)}$ are traceless, the same
must be true for their correlations. This implies their decoupling:

\begin{propo}
\bea\label{Arr}\merw{0}{A^{(r)}(e)(f),A^{(r')}(e')(f')} =
\delta_{rr'}N_r \cdot(-1)^r 
\sumno_{2n\leq r} \gamma^r_n\,
(E_{ff})^n (E_{f'f'})^n (E_{ff'})^{r-2n}
\eea
with the same coefficients $\gamma^r_n=
\frac{1}{4^nn!}\frac{r!}{(r-2n)!}\frac{1}{(1-r)_n}$ as in 
\eref{gamma}. The proper normalization 
$N_r=1$ can be achieved by adjusting $\alpha_r=\alpha^r_0$. 
\end{propo}

Proof: We make a general ansatz with coefficients $\gamma^{rr'}_{nn'}$ with
$r-2n=r'-2n'$. The vanishing of $\square_f$ and of $\square_{f'}$
gives conflicting recursions for $\gamma^{rr'}_{nn'}$ unless $r=r'$. If
$r=r'$, the recursion implies the displayed coefficients. \qed

While \eref{Ar} are defined for $m\geq0$, the
decoupling is exact only at $m=0$.

\begin{coro} \label{c:tlsymmc}
The massless symmetric tensor potentials $A^{(r)}(x,e)$ are traceless (by
construction) and conserved. They satisfy in addition the axial gauge
condition 
$$e^{\mu}A^{(r)}_{\mu\mu_2\dots\mu_r}(x,e)=0.$$
They are string-localized potentials given by the same
  formula \eref{as} (with $s$ replaced by $r\leq s$) for the massless
field strengths associated with the Wigner representations of helicity
$h=\pm r$ \cite{W}. They coincide with the potentials given in \cite[Sect.~4A]{PY}.
\end{coro}

Proof: When the divergence is taken, the derivative may be contracted
with an index of $E$ or with an index of
$a^{(r-2k)}$. The former contributions are $Ep=O(m^2)$, the
latter are $O(m)$ by \eref{eoms}, hence the divergence vanishes at
$m=0$. The axial gauge is a consequence of \cref{c:ea=0} and the fact
that $e^\mu E(e,e)_{\mu\nu}=0$. The last statements are immediate
because $E_{\mu\nu}$ differs from $\eta_{\mu\nu}$ by derivative terms
that do not contribute to the field strengths; and 
the coefficients are the same as in \eref{tpfK}.  
\qed

It remains to relate the normalization $N_r$ in \eref{Arr} 
(which should be $=1$ in the standard normalization \sref{s:m=0}) to
$\alpha_r=\alpha^r_0$ from \eref{Ar}. Because it is the coefficient of the
purely mixed term $(E_{ff'})^r$ in \eref{arr'}, it is easy to see from
\eref{Ar} and
\eref{Arr} that $N_r\gamma^r_0 = (\alpha_r)^2 \beta^{rr}_{00}$, with
$\beta^{rr}_{00} = \binom sr\inv\sum_{2m\leq s-r}
\frac{1}{4^mm!}\frac{(r-s)_{2m}}{(\frac12-s)_m}$ given by
\eref{brr'}. So the proper normalization is fixed by 
\bea\label{normal}(\alpha_r)^2=(\beta^{rr}_{00})\inv = \binom sr
\frac{\Gamma(\frac12+s)\Gamma(1+r)}{\Gamma(\frac12+\frac{r+s}2)\Gamma(1+\frac{r+s}2)}. 
\eea

\begin{remark}\label{r:m=0}
(i) The decoupled massless fields $A^{(r)}$ are independent of the spin
$s\geq r$ of the massive field in whose decomposition they emerge in
the massless limit. \\[1mm]
(ii) The axial gauge condition in \cref{c:tlsymmc} ensures the
reduction of the degrees of freedom as compared to the massive
representation of spin $r$ (relevant little group
$\SO(D-2)=\Eu(D-2)/\RR^{D-2}$ vs.\ $\SO(D-1)$ for $m>0$ in $D$ dimensions). \\[1mm]
(iii) For the \tpf s of the components $A^{(r)}_{\mu_1\dots\mu_r}$, the
factors $E_{ff}$, $E_{ff'}$ etc.\ in \eref{Arr} have to be replaced by
corresponding components of the tensors $E(e,e)(p)$:
$$\merw{0}{A^{(r)}_{\mu_1\dots\mu_r}(e),A^{(r)}_{\nu_1\dots\nu_r}(e')} =
(-1)^r\sum \wt {\gamma^r_n} \, (E(e,e)_{\mu\mu})^n (E(e',e')_{\nu\nu})^n(E(e,e')_{\mu\nu})^{r-2n}
$$
(notation as in \eref{tpf}). \\[1mm]
(iv) Taking the total curl, kills all factors
$p_\mu$ in all $E$ tensors. Therefore the \tpf s of the highest
field strengths $F^{(r)}_{[\mu_1\nu_1]\dots[\mu_r\nu_r]}$ are the same
as if they were derived from point-localized potentials $\AK{}^{(r)}$
with indefinite \tpf s \eref{tpfK}. These Feynman gauge potentials are
neither traceless nor conserved. \\[1mm]
\end{remark}

\begin{coro}\label{c:m=0}
The massless field strengths are independent of $e$, hence they
are point-localized fields, and 
\bea\label{AIF} A^{(r)}_{\mu_1\dots\mu_r}(x,e) = \big(I_e^r
F^{(r)}_{[\mu_1\nu_1]\dots[\mu_r\nu_r]}\big)(x)
e^{\nu_1}\dots e^{\nu_r}.
\eea
The formula \eref{e-dep} holds in the same way for the massless
potentials $A^{(r)}_{\mu_1\dots\mu_r}$.
\end{coro}

Proof: The first statement is (iv) of \rref{r:m=0}. \eref{AIF} follows
by the same argument as the one
leading to \eref{ass}. \eref{e-dep} for the massless potentials
$A^{(r)}_{\mu_1\dots\mu_r}(x,e)$ follows by the same argument as in
\cref{c:e-dep}. \qed

\cref{c:m=0} secures the string-independence of
massless actions $S_{\rm int}(e)$ like \eref{Se0}.

\subsection{``Fattening''}

The \tpf\ \eref{Arr} with $r=s$ is exact also for $m>0$.\footnote{This
  is not true for 
  the massive fields $A^{(r)}$ with $r<s$. Due to their coupling
  to fields with $r'>r$, their \tpf s are not just polynomials in
  $E_{\mu\nu}(p)$, cf.\ \eref{arr'}.}  
 
Thus, if one takes the massless string-localized potential $A^{(s)}\vert_{m=0}$ 
with \tpf\ \eref{Arr} (with $r=s$) as the starting point, one can 
get the mass by simply changing the dispersion relation
$p^0=\omega_m(\vec p)$ and taking the arguments of the functions
$E_{\mu\nu}(p)$ on the mass-shell. The previous analysis, where we
have derived this massive \tpf\ from a positive theory, shows that
this deformation preserves positivity. The fattened field
  brings along with it all lower rank fields $A^{(r)}$ by virtue of
  the coupling through the divergence, and the Proca field
  $\AP$ can be restored from the massive field $A^{(s)}$. This is
  possible because the deformation decreases the number of
null states of the \tpf, viewed as a quadratic form. Indeed, the
massive potential is not conserved, and hence it can create more
one-particle states.

\begin{remark} \label{r:fat}
The fattening allows to continuously
  ``turn on the mass'' in interactions with vector or tensor bosons
  without appealing to the Higgs mechanism and the ``eating of the
  Goldstone boson''. See the comments in \sref{s:overview}. 
\end{remark}

One can also get back the Proca potential $\AP(x)$ as
derivatives of the fattened potential $A^{(s)}(x,e)$:

\begin{propo}\label{p:fat} The point-localized Proca potential can be
  restored from the string-localized massive helicity $h=\pm s$ field
  $A^{(s)}\vert_{m>0}$ by ``applying the Proca \tpf\ \eref{tpf}'', regarded as a differential operator
($\pi_{\mu\nu}=\eta_{\mu\nu} +
m^{-2}\partial_\mu\partial_\nu$):
$$\AP_{\mu_1\dots\mu_s}(x) = (-1)^s\cdot \merw{m}
{\AP_{\mu_1\dots\mu_s},\AP{}^{\nu_1\dots\nu_s}} \cdot A^{(s)}_{\nu_1\dots\nu_s}\vert_m(x,e)$$
\end{propo} 

Proof: We multiply the \tpf\ in the form \eref{tpf} on $A^{(s)}$ in the
form \eref{Ar}. In the first step, we notice that every factor
$E_{\nu\nu}$ contained the field\footnote{\label{f:supp}We suppress sub-indices
    like $E_{\nu_i\nu_j}$ in this and all similar arguments to follow.} 
annihilates the \tpf\ because the latter is conserved and
traceless. Thus, we may replace $A^{(s)}$ by its leading term 
$a^{(s)}$ ($k=0$ in \eref{Ar}, $\alpha^s_0=\alpha_s=1$). In the second step, we notice that every factor
$\pi^{\nu\nu}$ in the \tpf\ annihilates $a^{(s)}$ by virtue of
\eref{eoms}. Thus we may replace the \tpf\ by its leading term $n=0$
in \eref{tpf}, which is $(-\pi)^{\otimes s}$. The claim then follows
from \eref{APA}.
\qed

\begin{propo}\label{p:invfat} Conversely, we have the formulae 
\bea\label{invfat1}a^{(s)}_{\mu_1\dots\mu_s}(x,e) = (-1)^s\cdot \merw{m}
{a^{(s)}_{\mu_1\dots\mu_s}(-e),a^{(s)\nu_1\dots\nu_s}(e)} \cdot
\AP_{\nu_1\dots\nu_s}(x)
\eea
for $m> 0$, and (after taking the limit $m\to0$ of the regular field $a^{(s)}$)
\bea\label{invfat2}A^{(s)}_{\mu_1\dots\mu_s}(x,e) = (-1)^s\cdot \merw{0}
{A^{(s)}_{\mu_1\dots\mu_s}(e),A^{(s)\nu_1\dots\nu_s}(e)} \cdot
a^{(s)}_{\nu_1\dots\nu_s}(x,e)
\eea
for $m=0$, to restore the massless helicity field $A^{(s)}$ from the
Proca field. In position space, the \tpf s \eref{ass}, \eref{Arr} are understood as integro-differential operators, cf.\ \eref{idiff}.
\end{propo}

\newpage

Proof: For \eref{invfat1}, we notice that every factor
$E^{\nu\nu}$ annihilates $\AP$ (traceless and conserved), hence
only $n=0$ in the \tpf\ contributes, and the factors $E_\mu{}^\nu$ act
on $\AP$ like $\delta_\mu^\nu-\frac{p_\mu 
  e^\nu}{(pe)_+}= J_\mu{}^\nu$. This gives $a^{(s)}$ by \eref{aru}. For
\eref{invfat2}, we notice that at $m=0$, $E_\mu{}^\nu$ acts on $a^{(s)}$ like
$\delta_\mu^\nu$ by the first of \eref{eoms} and \cref{c:ea=0}, and
$E^{\nu\nu}$ acts like $\eta^{\nu\nu}$. Thus, the second of
\eref{eoms} implies the claim. \qed

\section{Stress-energy tensor}
\label{s:SET}
\setcounter{equation}{0}
\subsection{The point-localized \set\ for $m>0$}

We refer to \aref{a:SET} for some comments on 
\set s and Lagrangians for free fields of higher spin. 

For our purposes here, it suffices to ``read back'' a suitable \set\
for the Proca field $\AP_{\mu_1\dots\mu_s}$ from a simple form of the
Poincar\'e generators. 

\begin{propo}\label{p:generators}
The generators of the Poincar\'e transformations of the Proca field
can be written as 
\bea\label{P}
P_\sig &\!=\!& (-1)^s \int d^3\vec x \Big[-\frac14
\AP_{\mu_1\dots\mu_s}\stackrel\lra{\partial_\sig}\stackrel\lra{\partial_0}
\AP{}^{\mu_1\dots\mu_s}\Big] , 
\\
\label{M} \hskip-2mm
M_{\sig\tau} &\!=\!& (-1)^s\int d^3\vec x\Big[-\frac{1}4\Big(x_\sig \cdot  
\AP_{\mu\times}\stackrel\lra{\partial_0}\stackrel\lra{\partial_\tau} \AP{}^{\mu\times}-(\sig\lra\tau)\Big)
- s 
\,\AP_{\sig\times}\stackrel{\lra}{\partial_0}\AP_\tau{}^\times\Big],\qquad
\eea
where $X_\times Y^\times$ stands for the contraction in $s-1$ indices
$\mu_2\dots\mu_s$.  
\end{propo}
Here and everywhere below, normal ordering is understood.

Before we give the proof, we state the corollary:
\begin{coro}\label{c:redset}
The generators \eref{P} and \eref{M} can be obtained from the
``reduced \set''
\bea\label{redset}
T^\red_{\rho\sig}:= (-1)^{s}\Big[-\frac{1}4
\AP_{\mu\times}\stackrel\lra{\partial_\rho}\stackrel\lra{\partial_\sig}
\AP{}^{\mu\times} - \frac s2\,\partial^\mu\Big(\AP_{\rho\times}\stackrel{\lra}{\partial_\sig} \AP_\mu{}^{\times}+(\rho\lra\sig)\Big)\Big].
\eea
\end{coro}
See \aref{a:SET} for how $T^\red$ relates to more familiar \set s. 

\eref{P} and the first term in \eref{redset} already appear in
\cite{Fz}. The second term in \eref{redset} does not contribute to the
momenta, but it produces the last term in \eref{M}, which is necessary in order to get the
correct infinitesimal boosts. This will become apparent in the proof
of \pref{p:generators}. 
The first term in \eref{redset} and the two parts of the derivative term are
separately conserved w.r.t.\ both indices $\rho$ and $\sig$ by
virtue of \lref{l:trivial}(i) resp.\ (ii). 

Proof of \cref{c:redset}: We have to do the integrals
\eref{generators} at fixed
$x^0=t$. The first part of \eref{redset} obviously gives \eref{P} and
the first terms of \eref{M}. The two pieces of the second part do not
contribute to $P_\sig$, and they give rise to the  
last term of \eref{M} by \lref{l:trivial}(i) and (ii), respectively. \qed

Proof of \pref{p:generators}: The argument for $P_\sig$ can
  essentially be found in \cite{Fz}, except that the commutator
  \eref{commrel} has been guessed not quite correct
  \cite[Eq.~(4.2)]{Fz}. We display the argument here because we shall
  use many variants of it below. See also fref{f:supp}.

The \tpf\ \eref{tpf} fixes the commutation relation 
\bea\label{commrel}
[\AP_{\mu_1\dots\mu_s}(x),\AP_{\nu_1\dots\nu_s}(y)] =
(-1)^sD_{\mu_1\dots\mu_s,\nu_1\dots\nu_s} \Delta_m(x-y) 
\eea
where 
$(-1)^sD_{\mu_1\dots\mu_s,\nu_1\dots\nu_s}=\merw{m}{\AP_{\mu_1\dots\mu_s},\AP_{\nu_1\dots\nu_s}}$
is the \tpf\ regarded as a differential operator
($\pi_{\mu\nu}=\eta_{\mu\nu} +
m^{-2}\partial_\mu\partial_\nu$) acting on the commutator function
$\Delta_m(x-y)$ of the scalar free field. The commutator of
$P_\sig$ with $\AP_{\nu_1\dots\nu_s}$ is 
$$ [P_\sig,\AP_{\nu_1\dots\nu_s}(y)]= -\frac12\int d^3\vec x\, D_{\mu_1\dots\mu_s,\nu_1\dots\nu_s}\Delta_m(x-y)\stackrel{\lra}{\partial_0}\stackrel{\lra}{\partial_\sig}\AP{}^{\mu_1\dots\mu_s}(x).$$ 
The derivatives $\partial_\mu$ appearing in pieces of the differential operator
$D$ can be partially integrated using \lref{l:trivial}(i), with 
$\Theta_{\rho\sig}$ of the form $\partial_\mu(D'_{\dots}\Delta_m\stackrel{\lra}{\partial_\rho}\stackrel{\lra}{\partial_\sig}\AP{}^{\mu\dots})$, 
suppressing further indices. 
After partial integration, the derivatives act on the field
$\AP{}^{\mu\dots}(x)$ where they vanish. Thus, one may replace
all operators of the form $\pi_{\mu\nu}$ and $\pi_{\mu\mu}$ in $D$ by
$\eta_{\mu\nu}$ and $\eta_{\mu\mu}$. Because the latter also kill the field
$\AP{}^{\mu\dots\mu}(x)$, only the contribution $n=0$ of the \tpf\
\eref{tpf} (that specifies the operator $D$) survives, and $D$ may be
replaced by the ``identity operator'' $(\eta_{\mu\nu})^{\otimes s}$. At this point, the integral can be
immediately performed: because \eref{P} integrated at $x^0=t$ is
independent of $t$, one may choose $x^0=y^0$, and use the
equal-time properties of the scalar commutator function:
$\Delta_m(x)\vert_{x^0=0}=0$ and
$\partial_0\Delta_m(x)\vert_{x^0=0}= -i\delta(\vec x)$. We get the
desired result $[P_\sig,\AP_{\nu\dots\nu}(y)]=-i\partial_\sig
\AP_{\nu\dots\nu}(y)$. 

The argument for the Lorentz generators is more involved. The commutator
of the first terms in \eref{M} with $\AP_{\nu_1\dots\nu_s}$ is
$$-\frac12\int d^3\vec x\, x_\sig D_{\mu_1\dots\mu_s,\nu_1\dots\nu_s}\Delta_m(x-y)
\stackrel\lra{\partial_0}\stackrel\lra{\partial_\tau}
\AP{}^{\mu_1\dots\mu_s}(x)-(\sig\lra\tau). $$
All terms involving $\partial_\mu\partial_\mu$, either from
$\pi_{\mu\mu}$ or from $\pi_{\mu\nu}\pi_{\mu\nu}$ within $D$, vanish
 because
$\int d^3\vec x\,\partial_\mu\big[ D''_{\dots} \Delta_m
\stackrel\lra{\partial_0}\stackrel\lra{\partial_\sig}\AP_\tau{}^{\mu\dots}\big]=0$
(using \lref{l:trivial}(i) twice).
Thus, the only contributions are due to
$(\eta_{\mu\nu})^{\otimes s}$ and $s$ terms
$(\eta_{\mu\nu})^{\otimes s-1}m^{-2}\partial_\mu\partial_\nu$. The former give
rise, if evaluated at $x^0=y^0$,
to the infinitesimal transformation of the point $x$:
$$-\frac12\int d^3\vec x\,
x_\sig\Delta_m(x-y)\stackrel\lra{\partial_0}\stackrel\lra{\partial_\tau}\AP_{\nu_1\dots\nu_s}
= -i (x_\sig\partial_\tau-x_\tau\partial_\sig)\AP_{\nu_1\dots\nu_s}.$$
The latter give rise, again by \lref{l:trivial}(i), to the undesired term 
$$-\frac1{2m^2} \sumno_{i=1}^s\int d^3\vec x\,\partial_{\nu_i} \Delta_m
\stackrel\lra{\partial_0}\stackrel\lra{\partial_\sig}\AP_{\tau\nu_1\dots\wh{\nu_i}\dots\nu_s}
- (\sig\lra\tau) = \frac i{m^2}
\sumno_{i=1}^s \partial_{\nu_i}\FP_{[\sig\tau]\nu_1\dots\wh{\nu_i}\dots\nu_s}.
$$

On the other hand, the commutator of the last term in \eref{M} with
$\AP_{\nu_1\dots\nu_s}$ is 
$$-s\int d^3\vec x\,D_{\sig\mu_2\dots\mu_s,\nu_1\dots\nu_s}\Delta_m(x-y)
\stackrel\lra{\partial_0}
\AP{}_\tau^{\mu_2\dots\mu_s}(x)-(\sig\lra\tau). $$
Again, all terms involving $\partial_\mu$ vanish by \lref{l:trivial}(i),
and terms involving $\eta_{\mu\mu}$ vanish because $\AP$ is
traceless. Thus, only the terms $\pi_{\sig\nu}(\eta_{\mu\nu})^{\otimes s-1}$ survive: 
$$=-\sumno_{i=1}^s\int d^3\vec x\,\big(\eta_{\sig\nu_i}+m^{-2}\partial_\sig\partial_{\nu_i}\big)\Delta_m(x-y)
\stackrel\lra{\partial_0}
\AP{}_{\tau\nu_1\dots\wh{\nu_i}\dots\nu_s}(x)-(\sig\lra\tau).$$
The contribution from $\eta_{\sig\nu_i}$ gives the infinitesimal
transformation of the tensor indices
$$-i\sumno_{i=1}^s\big(\eta_{\sig\nu_i} \stackrel\lra{\partial_0}
\AP{}_{\tau\nu_1\dots\wh{\nu_i}\dots\nu_s}-\eta_{\tau\nu_i} \stackrel\lra{\partial_0}
\AP{}_{\sig\nu_1\dots\wh{\nu_i}\dots\nu_s}\big).$$ 
The remaining contribution from $m^{-2}\partial_\sig\partial_{\nu_i}$ is
$$-\frac 1{m^2}\sumno_{i=1}^s\int d^3\vec x\,\partial_\sig\partial_{\nu_i}\Delta_m(x-y)
\stackrel\lra{\partial_0}
\AP{}_{\tau\nu_1\dots\wh{\nu_i}\dots\nu_s}(x)-(\sig\lra\tau)$$
and cancels with the previous undesired term thanks to the identity 
\bea\label{cancel} 
\int d^3\vec x\,\Big[
X\stackrel{\lra}{\partial_0}\stackrel{\lra}{\partial_\sig}Y +
2 \partial_\sig X\stackrel{\lra}{\partial_0}Y\Big] = \int d^3\vec
x\,\partial_\sig \Big[X\stackrel{\lra}{\partial_0}Y\Big]=0
\eea
(once more by \lref{l:trivial}(i), writing $\partial_\sig=\partial^\mu\eta_{\mu\sig}$).
\qed

\subsection{The string-localized stress-energy tensors for $m=0$}

We are going to separate ``irrelevant contributions'' from the reduced
\set, that do not contribute to the generators. It is, however, more
practical, to perform the corresponding partial integrations inside
the generators \eref{P}, \eref{M}, and read back a resulting \set, as
we have done before. In the first step, the partial integrations
remove all terms that are singular in the massless limit. 

We insert the preliminary decomposition \eref{exps} of the
point-localized potential $\AP$ in terms of derivatives of string-localized
fields $a^{(r)}$ into the Poincar\'e generators \eref{P} and
\eref{M}, and partially integrate all the derivatives of the
decomposition. The result is 
\begin{propo}\label{p:interm} Expressed in terms of string-localized
  fields $a^{(r)}$ ($r\leq s$), the Poin\-car\'e generators are
\bea\notag
P_\sig &\!=\!& \sumno_{r=0}^s\binom sr(-1)^r\, \int d^3\vec x \,\Big[-\frac14 
a^{(r)}_{\mu_1\dots\mu_r}(x,e)\stackrel\lra{\partial_0}\stackrel\lra{\partial_\sig}
a^{(r)\mu_1\dots\mu_r}(x,e')\Big],
\\ \notag
M_{\sig\tau} &\!=\!& \sumno_{r=0}^s\binom sr(-1)^r \, \int
d^3\vec x \\ \notag && \Big[-\frac{1}4\, x_\sig \, 
a^{(r)}_{\mu\times}(x,e)\stackrel\lra{\partial_0}\stackrel\lra{\partial_\tau}
a^{(r)\mu\times}(x,e') - \frac r2 
\,a^{(r)}_{\sig\times}(x,e)\stackrel{\lra}{\partial_0}a^{(r)}_{\;\;\tau}{}^\times(x,e')\Big]-(\sig\lra\tau)
\eea
for any pair $e$, $e'$, and at all values of the mass $m$. 
\end{propo}
\begin{remark}\label{r:cav}All
  quadratic expressions are understood as Wick products. As noticed in
 \cite{M}, under the Wick ordering the strings $e,e'$ may be set
 equal. We retain them to be independent, because this enlarges the
 class of \set s. 
\end{remark}

Proof of \pref{p:interm}: We insert the expansion \eref{exps} of
$\AP(x)$ in terms of derivatives of $a^{(r)}(e)$ 
resp.\ $a^{(r')}(e')$ into \eref{P}.  

It is routine work to partially integrate all the derivatives coming
from \eref{exps}, using \lref{l:trivial}(i) again and again. The field
equations \eref{eoms} produce positive powers of the mass $m$, that
cancel all inverse powers of the expansion: Partially integrating  
$\partial_{\mu}a^{(r)}_{\dots}(e)$ against $a^{(r')}{}^{\mu\dots}(e')$, 
one gets $m a^{(r)}_{\dots}(e) \cdots
a^{(r'-1)}{}^{\dots}(e')$ by \eref{eoms}, and vice 
versa. Partially integrating $\partial_\mu a^{(r)}_{\dots}(e)$
against $\partial^\mu a^{(r')}{}^{\dots}(e')$, one gets $m^2
a^{(r)}_{\dots}(e)\cdots a^{(r')}{}^{\dots}(e')$ by the
Klein-Gordon equation. In the expansion of the momenta \eref{P}, the
number of terms with $a$ contractions between derivatives, $b$
contractions between $a^{(r)}(e)$ and a derivative, $b'$
contractions between a derivative and $a^{(r')}(e')$, and $c$
contractions between $a^{(r)}(e)$ and 
$a^{(r')}(e')$, such that $r=b+c$, $r'=b'+c$ and $a+b+b'+c=s$, is 
$\frac{s!}{a!b!b'!c!}$.  Each such term after partial integration
becomes (schematically) $(-1)^{b+b'} a^{(c)} \cdots
a^{(c)}$ times the same operator quadratic in $a^{(c)}$. Therefore the
combinatorics is done by observing that  
$\sum_{a+b+b'=s-c}(-1)^{b+b'}\frac{s!}{a!b!b'!c!} = (-1)^{s-c}\binom sc$. 

The expansion of the Lorentz generators \eref{M} is likewise just a
counting issue, where special care has to be taken with the
tensor indices $\sig$, $\tau$ in the second contribution to
$M_{\sig\tau}$. When they are attached to derivatives, they cancel
against the results of partial integrations according to
\lref{l:trivial}(i) in the first term.\qed  

The formulae in \pref{p:interm} have the merit that they do not
contain any singular fields, and one may read back a conserved and
symmetric massive string-localized \set\ $T^{\rm reg}_{\sig\rho}(e,e')$
that is regular at $m=0$, in exactly the same way as was done in \cref{c:redset}
from \pref{p:generators}. The limit $m\to 0$ can be taken directly
by putting $m=0$. But these steps are of little use,
because the intermediate escort fields $a^{(r)}$ do not decouple. 
We must in turn express $a^{(r)}$ in \pref{p:interm} in terms of the
decoupling string-localized fields $A^{(r-2k)}$. The following result
holds only at $m=0$, where the decoupling of \tpf s is exact.  

\begin{propo}\label{p:decoup}
At $m=0$, one has 
\bea\label{decoup}
P_\sig = \bigoplus\nolimits_{r=0}^s P^{(r)}_\sig,\qquad M_{\sig\tau} =
\bigoplus\nolimits_{r=0}^s M^{(r)}_{\sig\tau}
\eea
where for any $e,e'$ 
\bea\label{Psl}
P^{(r)}_\sig &\!\!=\!\!& (-1)^r \,\int d^3\vec x \, \Big[-\frac14 
A^{(r)}_{\mu_1\dots\mu_r}(x,e)\stackrel\lra{\partial_0}\stackrel\lra{\partial_\sig}
A^{(r)\mu_1\dots\mu_r}(x,e')\Big],
\\\label{Msl} 
M^{(r)}_{\sig\tau} &\!\!=\!\!& (-1)^{r}\,\int d^3\vec x\\ \notag && \hskip-10mm
\Big[-\frac14\,x_\sig \,  
A^{(r)}_{\mu\times}(x,e)\stackrel\lra{\partial_0}\stackrel\lra{\partial_\tau}
A^{(r)\mu\times}(x,e') -\frac r2
\,A^{(r)}_{\sig\times}(x,e)\stackrel{\lra}{\partial_0}A^{(r)}_{\tau}{}^\times(x,e')
\Big] -(\sig\lra\tau).
\qquad
\eea
The notation in \eref{decoup} asserts that the generators
$P_\sig^{(r)}$ and $M_{\sig\tau}^{(r)}$ commute with $A^{(r')}$ and
consequently with $P_\sig^{(r')}$ and $M_{\sig\tau}^{(r')}$
($r'\neq r$), and hence generate the infinitesimal Poincar\'e
transformations of $A^{(r)}$ according to \eref{covariance}.
\end{propo}

Proof: We insert the expansion \eref{Ar} in terms of
$E_{\mu\mu}(e,e)^k a^{(r-2k)}_{\mu\dots\mu}(e)$ into $A^{(r)}(e)$
in \eref{Psl}. We partially integrate the derivatives contained in the
factors $E(e,e)$ (cf.\ \eref{idiff}). When they hit $A^{(r)}(e')$, they vanish because $A^{(r)}$ are conserved at $m=0$. The 
remaining contribution $\eta_{\mu\mu}$ of $E_{\mu\mu}$ is directly contracted
with $A^{(r)}(e')$, and vanishes because $A^{(r)}$ are
traceless at $m=0$. Thus, only the leading term
$A^{(r)}(e)=\alpha_ra^{(r)}(e) + \dots$ contributes. Now,
we expand $A^{(r)}(e')$ and partially integrate the
derivatives contained in $E^{\mu\mu}(e',e')$ onto
$a^{(r)}(e)$, where they vanish because $a^{(r)}$ are conserved
at $m=0$. But $a^{(r)}$ are not traceless, and $E(e',e')^k$ acts like
$\eta^ka^{(r)}(e)=(-1)^ka^{(r-2k)}(e)$ by \eref{eoms}. It
remains to add up the coefficients  
$$\sumno_{2k\leq s-r}(\alpha_{r+2k})^2 \gamma^{r+2k}_k = \binom sr.$$
(We were not able to establish this identity for finite sums of
rational numbers in closed form, but have verified it numerically
until $s=100$.) 

Again, the case of the Lorentz generators requires a more involved
combinatorics. Let us consider the first step: the partial 
integration of derivatives $\partial_\mu a'(e)$ contained in
$E(e,e)^k a^{(r-2k)}(e)$ against $A^{(r)}(e')$. By 
\lref{l:trivial}(i), the partial integrations within the first term in
\eref{Msl} give undesired non-vanishing contributions of the form 
$$-\frac14\cdot
2\cdot\frac{r(r-1)}2\int d^3\vec x\, \Big[a'(x,e)\stackrel{\lra}{\partial_0}\stackrel{\lra}{\partial_\sig}A^{(r)}_{\tau}{}^{\dots}(x,e')\Big]-(\sig\lra\tau),$$
where the factor $2\cdot\frac{r(r-1)}2$ counts the assignments
of the other contracted indices. On the other hand, when the index
$\sig$ is attached to a factor $E$ 
in the second term of \eref{Msl}, it gives the undesired term
$$-\frac r2\cdot (r-1)\int d^3\vec x\, \Big[\partial_\sig
a'(x,e)\stackrel{\lra}{\partial_0}A^{(r)}_{\tau}{}^{\dots}(x,e')\Big]-(\sig\lra\tau)$$
with another counting factor. These terms cancel each other by virtue
of \eref{cancel}.
In the second step: the partial
integration of derivatives $\partial^\mu a''(e')$ contained in
$E(e',e')$ within $A^{(r)}(e')$
against $a^{(r)}(e)$, the
cancellations occur with the same pattern. This shows the equality of
the generators in \pref{p:decoup} and \pref{p:interm}. 

The final statements are immediate: 
$A^{(r)}$ mutually commute, because their mixed \tpf s vanish. Hence
the ``$r$'' generators commute with the ``$r'$'' fields and
generators. Then the ``$r$'' generators act on the ``$r$'' fields like
the full generators $P_\sig$ and $M_{\sig\tau}$, hence they implement the
correct Poincar\'e transformations. \qed

One can now read back conserved and symmetric massless string-localized
\set s $T^{(r)}_{\sig\rho}(r)$ from
\eref{Psl}, \eref{Msl}.

\begin{propo}\label{p:set} The generators \eref{Psl} and \eref{Msl}
  can be obtained from the string-localized massless \set s
  for every $r\geq 1$:
\bea\label{Tr}
T^{(r)}_{\rho\sig}(x,e,e')&:=& (-1)^{r}\Big[-\frac14
A^{(r)}_{\mu\times}(x,e)\stackrel\lra{\partial_\rho}\stackrel\lra{\partial_\sig}
A^{(r)}{}^{\mu\times}(x,e') \\ \notag && \qquad\quad -\, \frac r4\,\partial^\mu\Big(A^{(r)}_{\rho\times}(x,e)\stackrel{\lra}{\partial_\sig}
A^{(r)\times}_{\,\,\mu}(x,e')\ba{c}+(e\lra e')\\+(\rho\lra\sig)\ea\Big)\Big].\quad
\eea
\end{propo}
Proof: The argument is the same as with \cref{c:redset}. \qed

The \set s $T^{(r)}$ do not depend on the spin $s\geq r$ of the
reduced \set\ \eref{redset} from which they were extracted at $m=0$. 
By \eref{AIF}, they can also be expressed in terms of the
corresponding field strengths $F^{(r)}$, that are directly obtained
from the massless helicity $h=\pm r$ Wigner representations \cite{W}.

\begin{remark} For the charge operator
$$
Q=(-1)^si\int_{x^0=t} d^3\vec x \, \AP{}^{*}_{\mu_1\dots\mu_s}(x)
\stackrel{\lra}{\partial_0} \AP{}^{\mu_1\dots\mu_s}(x) 
$$
for complex potentials, one can proceed in complete analogy as
with the momentum operators, and obtains string-localized 
massless conserved currents
\bea
J^{(r)}_{\rho}(x,e,e') = (-1)^r \, \frac i2 \Big(
A^{(r)*}_{\mu_1\dots\mu_r}(x,e)\stackrel\lra{\partial_\rho}
A^{(r)}{}^{\mu_1\dots\mu_r}(x,e') + (e\lra e')\Big).
\eea
\end{remark}

The string-localized densities $T^{(r)}_{\rho\sig}$ and $J^{(r)}_\rho$
may be averaged over the directions of their strings (cf.\
\rref{r:cav}) with test functions of arbitrarily small support. Hence,
they can be localized in arbitrarily narrow spacelike cones.

\section{Conclusion}
We have introduced string-localized potentials for massive particles of
integer spin $s$, that admit a smooth massless limit to potentials with
individual helicities $h=\pm r$, $r\leq s$. We have elaborated several
remarkable properties of the massless limit, including an inverse
prescription how to pass from the massless to the massive potentials via a
manifestly positive deformation of the \tpf.

As a byproduct, we could construct string-localized currents and \set s for
massless fields of any helicity, that evade the Weinberg-Witten
theorem in a very conservative way.

Our results also allow to approximate string-localized fields in
  the massless infinite-spin Wigner representations \cite{MSY} by the
  massive scalar escort fields $A^{(0)}$ in the limit $s\to\infty$,
  $m^2s(s+1)=\kappa^2=$ const.\ \cite{R17}.

The feature of string-localization arises just by multiplication
operators in momentum space (of a special form), acting on the intertwiner
functions that define covariant fields in terms of creation and
annihilation operators of the $(m,s)$ Wigner representations.  

In particular, string-localization of the fields does not change the
nature of the particles that they describe, nor does it relax any of
the fundamental principles of relativistic quantum field theory. We
emphasize that we regard fields (associated with a given particle)
mainly as a device to formulate interaction Lagrangians. String-localized
interactions are admissible whenever their string-dependence is a
total derivative. In that case, string-localized fields have the
primary benefit of a better UV behaviour than point-localized fields
associated with the same particles. They therefore admit the
formulation of interactions that are otherwise only possible at the
expense of introducing states of negative norm and compensating ghost
fields. 

The renormalized perturbation theory of interactions mediated by
string-localized fields is presently investigated. It bears 
formal analogies with BRST renormalization, but is more economic (by
avoiding auxiliary unphysical degrees of freedom), and much closer to
the fundamental principles of relativistic quantum field theory. 

The necessity of using string-localized quantities to connect the
vacuum state with scattering states in theories with short-range
interactions was exhibited much earlier by Buchholz and Fredenhagen \cite{Bu,BF} 
investigated, in the framework of algebraic quantum field theory,
the localization properties of particle states in charged sectors relative to
the vacuum. Their conclusion was that, depending on the given model, the
best possible localization is in an arbitrarily narrow spacelike cone,
and that in the presence of a mass gap it cannot be worse in general.

The emerging renormalized perturbation theory using string-localized
fields \cite{S15,S16,M,MS,GMV} is the practical realization of this insight.

\bigskip

{\bf Acknowledgements.}
JM and KHR were partially supported by CNPq under grant No
312963/2013-0, JM also by FAPESP, CAPES and Finep. KHR and BS 
enjoyed the hospitality of the UF de Juiz de Fora, where parts of this
work were done. We thank D. Buchholz for pointing out ref.\
\cite{Lp}. We thank the referees for pointing out
  further references on higher spin theory and stimulating us to elaborate
  more comprehensively on the conceptual background.

\appendix

\section{Stress-energy tensors for higher spin fields}
\label{a:SET}
\setcounter{equation}{0}

\cite{FR} and \cite{GM} give excellent discussions of
how to properly define \set s. We focus only on a few facts. 

It is well-known from the example of the
free Maxwell field, that the canonical definition 
$$ T_{\rho\sig}= \sum\frac{\partial L}{\partial\partial^\rho\phi}\partial_\sig
\phi-\eta_{\rho\sig} L[\phi],$$
where the sum extends over all independent fields, may not give rise
to a symmetric \set. Consequently its Lorentz generators defined 
by \eref{generators} are not time-independent, even if $L$ is Lorentz
invariant. In the Maxwell case, the canonical \set\ is also not gauge
invariant, and both defects can be cured ``in one stroke'' by adding the
trivially conserved term $\partial_\kap(F^{\mu\kap}A_\nu)$. There are
other prescriptions (e.g., \cite{B,R}) to obtain symmetric \set
s in the general case.

The modern approach uses the Hilbert \set\ that is defined by varying
a generally covariant version of the action \cite{R,HE,FR,GM} with
respect to the metric, and then putting $g_{\mu\nu}=\eta_{\mu\nu}$:
\bea\label{Hilb}
T^{\rm (Hilbert)}_{\rho\sig}(x):=2\frac{\delta S}{\delta
  g^{\rho\sig}(x)}\Big\vert_{g=\eta}.
\eea
The Hilbert tensor is always symmetric and conserved. In both
approaches, one first needs a Lagrangian whose Euler-Lagrange
equations are the equation of motion.

This question has been addressed by Fierz and Pauli \cite{FP} 
and Fronsdal \cite{Fd} for free massive spin s fields; they used
auxiliary fields to ensure the vanishing of the divergence. When
varying with respect to the metric, one may omit terms involving the
divergence and the auxiliary fields that vanish by virtue of the
equations of motion. For $s=2$, this gives  
\bea\label{L2}
L'=\frac14 \FP_{[\mu\nu]\kap}\FP{}^{[\mu\nu]\kap} -\frac{m^2}2
\AP_{\nu\kap} \AP{}^{\nu\kap}.
\eea
The generally covariant action is
\bea\label{L2gen}
S=\int d^4x\,\sqrt{-g}\,\Big(\frac14 g^{\mu\mu'}g^{\nu\nu'}\FP_{[\mu\nu]\kap}\FP_{[\mu'\nu']\kap'} -\frac{m^2}2
g^{\nu\nu'}\AP_{\nu\kap} \AP_{\nu'\kap'}\Big) g^{\kap\kap'}
\eea
where 
$\FP_{[\mu\nu]\kap}:=D_\mu\AP_{\nu\kap}-D_\nu\AP_{\mu\kap}=\partial_\mu\AP_{\nu\kap}-\partial_\nu\AP_{\mu\kap}
-
(\Gamma^\lam_{\mu\kap}\AP_{\nu\lam}-\Gamma^\lam_{\nu\kap}\AP_{\mu\lam})$.  
The variation of $g^{\mu\mu'}$ and $g^{\nu\nu'}$ and the factor $\sqrt{-g}$ in $S$ give the \set\
\bea\label{Tfz}
T^\fz_{\rho\sig} = \eta^{\lam\lam'}\FP_{[\rho\lam]}{}^\mu\FP_{[\sig\lam']\mu} -m^2
\AP_\rho{}^{\mu} \AP_{\sig\mu}-\eta_{\rho\sig} L'
\eea
This tensor was first considered by Fierz \cite{Fz}. However, 
unlike the case of antisymmetrized indices, the Christoffel symbols
for the indices $\kap,\kap'$ do not drop out; and the contraction
by $g^{\kap\kap'}$
carries another dependence on the metric, so that we have
\begin{propo} The Hilbert \set\ is
$T^{\rm (Hilbert)}_{\rho\sig} = T^\fz_{\rho\sig} + \Delta T_{\rho\sig}$
with
\bea\label{DeltaT}
\Delta T_{\rho\sig} = -\frac12\,\partial^\mu\Big[\AP_\rho{}^\lam
\FP_{[\sig\lam]\mu}+\AP_\sig{}^\lam
\FP_{[\rho\lam]\mu}+\AP_\mu{}^\lam\big(\FP_{[\lam\rho]\sig}+\FP_{[\lam\sig]\rho}\big)\Big].
\eea
\end{propo}
Fierz \cite{Fz} has shown that $T^\fz$ produces the Hamiltonian 
$$P_0 = -\frac14 \int d^3\vec x \,
\AP{}^{\mu\nu}\stackrel\lra{\partial_0}\stackrel\lra{\partial_0}
\AP_{\mu\nu}, 
$$
and one easily verifies that the commutator is
$i[P_0,\AP_{\mu\nu}]=\partial_0\AP_{\mu\nu}$.\footnote{See
  \pref{p:generators}.} The same
is true for all $P_\sig$. Fierz has actually given a hierarchy of $s$
linearly independent \set s $T^{(q)}$ for the free massive spin $s$ field. 
They involve an increasing number $q=1,\dots,s$ of derivatives of the
potential, and overall factors $(-2m^2)^{-(q-1)}$. They all produce the
same generators $P_\sig$ that implement the correct infinitesimal translations
$i[P_\sig,\AP_{\mu_1\dots\mu_s}]=\partial_\sig\AP_{\mu_1\dots\mu_s}$. 

The Fierz \set s also all produce the same generators $M^\fz_{\sig\tau}$, 
but the latter do not implement the correct infinitesimal Lorentz
transformations! E.g., for $s=2$, one finds $i[M_{0i}^\fz,\AP_{00}] =
(x_0\partial_i-x_i\partial_0)\AP_{00} + A_{i0} - m^{-2}\partial_0
\FP_{[0i]0}$ rather than the correct $i[M_{0i},\AP_{00}] =
(x_0\partial_i-x_i\partial_0)\AP_{00} + 2A_{i0}$. 

This defect is precisely cured by the correction $\Delta T_{\rho\sig}$
in the Hilbert \set, given in \eref{DeltaT}. 

For spin $s>2$, the situation is worse. As
for $s=2$, the Fierz \set s do not generate the correct Lorentz
generators. The covariant generalizations of higher spin \set s
involving auxiliary fields \cite{Fd} suffer from inconsistencies, so
that their variation w.r.t.\ the metric is problematic. Nevertheless,
let us naively generalize \eref{L2}:\footnote{Much as \eref{L2}, this
  is not a valid Lagrangian since it does not entail the constraints.}
$$
L'=(-1)^s\Big[\frac14 \FP_{[\mu\nu]\kap_2\dots\kap_s}\FP{}^{[\mu\nu]\kap_2\dots\kap_s} -\frac{m^2}2
\AP_{\nu\kap_2\dots\kap_s} \AP{}^{\nu\kap_2\dots\kap_s}\Big],$$ 
make it generally covariant as in
\eref{L2gen}, and compute its Hilbert tensor. The result is $T^{\rm (Hilbert)}=T^\fz+\Delta T$ where $T^\fz$ 
is exactly as in \eref{Tfz} (all additional, un-curled indices
$\kappa_3\dots\kappa_s\equiv\times$ contracted) with the overall sign $(-1)^s$ (due to our
sign convention of the metric), and 
$$
\Delta T_{\rho\sig} = -(-1)^s\frac{s-1}2\,\partial^\mu\Big[\AP_\rho{}^{\lam\times}
\FP_{[\sig\lam]\mu\times}+\AP_\sig{}^{\lam\times}
\FP_{[\rho\lam]\mu\times}+\AP_\mu{}^{\lam\times}\big(\FP_{[\lam\rho]\sig\times}+\FP_{[\lam\sig]\rho\times}\big)\Big]
$$
arising from the variation of the metric and Christoffel symbols
associated with each of the $s-1$ contracted indices $\kappa_2\dots\kappa_s$ in $L'$. 
$T^{\rm (Hilbert)}$ differs from the reduced \set\ \eref{redset} only 
by ``irrelevant derivative terms'', 
in the sense that it gives the same Poincar\'e generators \eref{P} 
and \eref{M}. Clearly, it does not have a massless limit either, because of
the factors up to $m^{-2s}$ in $\merw{m}{\AP,\AP}$.

The prescription just outlined does not look satisfactory. 
  Indeed, our strategy of ``reading back'' a \set\ from the correct
  Poincar\'e generators, as we have done in \cref{c:redset}, is an
  alternative prescription that does not need a classical Lagrangian. 
  Because the ``correct generators'' are
  determined by their commutation relations with the field, which in
  turn are dictated by the Wigner representation theory, this approach
  is intrinsically quantum theoretic.

(We are not aware of a general argument that the Hilbert tensor
always, also in the presence of constraints, yields the correct generators.
This issue is not explicitly mentioned in the literature, including
the reviews \cite{GM,FR}.)

That $T^\red$ and $T^{\rm (Hilbert)}$ differ only by irrelevant
derivative terms, can be verified by hand (but we spare the reader
this cumbersome exercise). One may first rewrite $T^\fz$ with the help
of the identities 
$$\eta^{\lam\lam'}\FP_{[\rho\lam]\times}\FP_{[\sig\lam']}{}^{\times} -m^2
\AP_{\rho\times}\AP_\sig{}^\times = \FP_{[\rho\mu]\times}\partial_\sig
\AP{}^{\mu\times}-\partial^\mu \big(\FP_{[\rho\mu]\times}\AP_\sig{}^\times\big) $$
and
$$-\frac14
\FP{}^{[\mu\nu]\times}\FP_{[\mu\nu]\times}+\frac{m^2}2\AP{}^{\nu\times}\AP_{\nu\times}
= -\frac12\, \partial^\mu
\Big[\AP{}^{\nu\times}\FP_{[\mu\nu]\times}\Big],$$
then add $\Delta T$, and finally show that the difference from 
\eref{redset} does not contribute to the generators according to
\lref{l:trivial}(i) and (ii) (where $\Theta$ are various contributions to the \set).

\section{A useful lemma}
\label{a:lemma}
\setcounter{equation}{0}

The following (rather trivial, but very useful) lemma deals with a
covariant form of partial integration of four-derivatives in spatial
(fixed-time) integrals.  

\begin{lemma}\label{l:trivial}
With a tensor $\Theta_{\rho\sig}$ we associate the
``charges'' (not necessarily independent of $t$) $\Pi_\sig :=
\int_{x^0=t} d^3\vec x\, \Theta_{0\sig}$ and  
$\Omega_{\sig\tau} := \int_{x^0=t} d^3\vec x\, (x_\sig \,\Theta_{0\tau} - x_\tau\,
\Theta_{0\sig})$. We assume all fields or functions to have
  sufficiently rapid decay in spatial directions, so that boundary
  terms do not matter.

{\rm(i)} If $\Theta_{\rho\sig}$ is of the form 
$$\Theta_{\rho\sig} = \partial^\mu \big(Y_\mu
\stackrel\lra{\partial_\rho} Z_\sig\big)$$
(or a sum of terms\footnote{A term $Y
  \stackrel\lra{\partial_\rho} Z_{\mu\sig}=Y\delta_\mu^\lam
  \stackrel\lra{\partial_\rho} Z_{\lam\sig}$ can be written as a sum
  over terms of this form.} of the same structure), where $Y$ and $Z$ are
solutions to the Klein-Gordon equation, then
$\partial^\rho\Theta_{\rho\sig}=0$ trivially. The charges   
$\Pi_\sig := \int_{x^0=t} d^3\vec x\, \Theta_{0\sig}=0$
vanish, and the charges $\Omega_{\sig\tau}$ are 
$$\Omega_{\sig\tau} = \int_{x^0=t} d^3\vec x\, \big(Y_\tau
\stackrel\lra{\partial_0} Z_\sig - Y_0 \stackrel\lra{\partial_\tau}
Z_\sig\big) -(\sig\lra\tau).$$
{\rm(ii)} The same is true with 
$$\Theta_{\rho\sig} = \partial^\mu X_{[\mu\rho]\sig},$$
where $[\mu\rho]$ stands for an anti-symmetric index pair, and 
$$\Omega_{\sig\tau} = \int_{x^0=t} d^3\vec x\, \big( X_{[\tau 0]\sig} - X_{[\sig 0]\tau}\big).  $$
{\rm(iii)} In order for $\Omega_{\sig\tau}$ to vanish, the respective
integrands have to be spatial derivatives.
\end{lemma}

Proof: $\Theta_{0\sig}=\partial^\mu X_{[\mu 0]\sig}$ in (ii) is a spatial derivative, because
the term $\mu=0$ is absent by anti-symmetry. The claim follows by
partial integration. (i) is a special case of (ii) by writing 
$\Theta_{\rho\sig} = \partial^\mu \big(Y_\mu
\stackrel\lra{\partial_\rho} Z_\sig-Y_\rho
\stackrel\lra{\partial_\mu} Z_\sig\big)$.
The statement (iii) is trivial.
\qed

\end{document}